 \documentclass[aps,prc,twocolumn,linenumbers]{revtex4}
\usepackage{graphicx}
\usepackage{graphics}
\usepackage{amsmath,bbm}
\usepackage{amssymb,bm}
\usepackage{array}
\usepackage{longtable}
\usepackage{footnote}
\usepackage{xcolor}
\begin{document}

\title{Charmonium suppression in ultra-relativistic Proton-Proton collisions at LHC energies: A hint for QGP in small systems}
\author{Captain R. Singh$^{1}$}
\email[]{captainriturajsingh@gmail.com}
\author{Suman Deb$^{1}$}
\email[]{Suman.Deb@cern.ch}
\author{Raghunath Sahoo$^{1,2}$}
\email[]{Raghunath.Sahoo@cern.ch (Corresponding ~Author)}
\author{Jan-e Alam$^{3}$}
\email[]{jane@vecc.gov.in}
\affiliation{$^{1}$Department of Physics, Indian Institute of Technology Indore, Simrol, Indore 453552, India}
\affiliation{$^{2}$CERN, CH 1211, Geneva 23, Switzerland}
\affiliation{$^{3}$Variable Energy Cyclotron Centre, 1/AF, Bidhan Nagar, Kolkata—700064, India}

\begin{abstract}
Proton-proton ($pp$) collision has been considered as a baseline to study the system produced in relativistic heavy-ion (AA) collisions with the basic assumption that no thermal medium is formed in $pp$ collisions. This warrants a cautious analysis of the system produced in $pp$ collisions at relativistic energies.In this work we investigate the charmonium suppression in $pp$ collisions at $\sqrt{s} = 5.02, 7$ and $13$ TeV energies. Further, charmonium suppression has been studied for various event multiplicities and transverse momenta by including the mechanisms of color screening, gluonic dissociation, collisional damping along with regeneration due to correlated $c\bar c$ pairs. Here we obtain a net suppression of charmonia at high-multiplicity events indicating the possibility of the formation of quark-gluon plasma in $pp$ collisions.
\end{abstract}
\maketitle 
\section{Introduction}
\noindent

The heavy-ion programs at the Relativistic Heavy Ion Collider (RHIC) and Large Hadron Collider (LHC) are dedicated to understanding the rich and complex nature of the phase transition at high temperature and/or density from hadrons to a 
deconfined state of thermal quarks and gluons. This new state of deconfined
colored quarks and gluons is called quark-gluon plasma (QGP)~\cite{qgp1,qgp2,QWG}.  The suppression of quarkonia~\cite{matsui} has been proposed as a signal of the formation of the transient phase of QGP. 
The production of heavy quark pairs ($c\bar{c}$ or $b\bar{b}$) takes place at the initial stage of the collisions which can be treated perturbatively but the evolution of the quark pairs and the formation of their
bound states (quarkonia) occur through non-perturbative processes~\cite{bodwin1}. A quarkonia state traversing through QGP gets dissociated to unbound quark anti-quark pair, which results in the reduction of quarkonia yields (which will otherwise remain as a bound state) referred as quarkonia suppression. The quarkonia production gets altered by the cold nuclear matter (CNM) effects (nuclear absorption and shadowing) ~\cite{vogt}. The asymmetric $p$A collision can be used to disentangle the CNM effects from the QGP medium effects~\cite{phenix}. The $p$A collision has been used to serve as an important baseline 
for the understanding and interpretation of the heavy-ion (AA)  collision data. As the CNM effects will be present both in $p$A and AA collisions, and also 
it is expected that QGP will be formed in AA but not in $p$A collisions,
therefore, the data from $p$A may be used to distinguish the effects of QGP. In fact, the $p$A  experimental data corresponding to quarkonia suppression have been explained by considering CNM effects only at various rapidities and collision centralities. For instance, the suppression pattern of  $J/\psi$ in d$-$Au collisions at RHIC is well explained by the CNM effects \cite{PHENIX:2007tnc}.\\ 

It is pertinent to mention here that the formation of hot QGP medium in AA collisions has been predicted by considering the ratio of hadronic spectra originating from AA to $pp$ collisions. While calculating this ratio, the spectra from $pp$ collisions required to be appropriately scaled up by the number of nucleon-nucleon collisions in AA interaction. The deviation of the ratio from unity has been treated as a signal of medium formation in AA interactions.
The data from $pp$ collision have been used in these studies as a benchmark. The formation of QGP and its collective nature has also been studied by evaluating various flow coefficients (like directed, elliptic, triangular, etc)~\cite{flow,flow1}. For example, the elliptic flow coefficient has been used to extract the ratio of shear viscosity ($\eta$)  to entropy density ($s$) and the triangular flow coefficients have been used to understand the initial-state fluctuations.\\

It was expected that the elliptic flow for small systems formed in $pp$ and $p$A collisions will be zero due to lack of thermalization in such systems. However, data from p$-$Pb collisions at $\sqrt{s_{NN}}\; = \; 5.02,\; \& \;8.16$ TeV and for $pp$ collisions at $\sqrt{s}\; = \; 7\; \& \;13$ TeV at the LHC have provided hints towards the formation of a QGP medium~\cite{ppbj,pPbB,alice2,alice3,atlas2,atlas3,cms2}. The search for QGP in p$-$Pb collisions at $\sqrt{s_{NN}}\; = \; 5.02$ TeV has been carried out by the  theoretical model which will be called, 
``Unified Model of Quarkonia Suppression (UMQS)'' henceforth. 
This model is based on the unification of the various suppression and regeneration mechanisms of quarkonia produced in ultra-relativistic collisions~\cite{capt3}. In the present work, UMQS model approach is employed to investigate the formation of QGP in $pp$ collisions through charmonium suppression. The dissociation temperature ($T_D$) of charmonia states is much less than in bottomonia states. Therefore, in this work, we choose charmonia over bottomonia because the temperature of the system formed in $pp$ collisions will not be high enough to dissociate bottomonia ($T_D>650$ MeV). In previous works, it is found that bottomonia hardly gets suppressed even in p$-$Pb collisions at $\sqrt{s_{NN}}\;=\; 5.02$ TeV \cite{capt3}. Therefore, charmonia is a better choice to investigate QGP formation in small systems.\\

The dissociation of the quarkonia  ($J/\psi,~\Upsilon$, etc.) 
occurs through various mechanisms~\cite{matsui, nendzig,capt,ganesh}. 
As mentioned, charmonium suppression in heavy-ion collisions can be  
caused by the cold nuclear matter (CNM) effects  as well as by the 
Debye screening of color interaction, gluonic dissociation 
and collisional damping in hot  QGP
~\cite{matsui,nendzig, wols, sharma,brambilla, peskin,bhanot}. 
As the $pp$ collisions lack the nuclear environment, {\it i.e.} the CNM effects, therefore,
the hot QGP effects are considered to study the  
charmonia suppression in $pp$ collision. 
Along with the suppression, there is a non-zero probability that a 
charmonia state can be regenerated in the medium
at temperature ($T$), $T<T_{D}$.\\

In the current study, we have used the UMQS model to predict the charmonia suppression in ultra-relativistic 
$pp$ collisions. This model has been used  to
explain the bottomonia suppression in Pb$-$Pb 
and p$-$Pb collisions at LHC energies~\cite{capt,capt3,nikhil}. UMQS includes 
color screening, gluonic dissociation, collisional damping, and regeneration as the hot matter 
effects. Due to the absence of a nuclear environment in $pp$ collisions, 
the cold nuclear matter 
effects like shadowing need not be considered here. 
We have used the  second-order dissipative fluid dynamics~\cite{amuronga,mis1,mis2,mis3}
in (1+1) dimension
with the assumption of boost invariance to study the propagation of the
charmonia in the expanding QGP background.\\

The paper is organized as follows.  The space-time evolution of QGP is discussed in section II.
Section III  is devoted for studying the charmonium kinematics and medium effects along with   
brief discussion of UMQS. The model prediction and comparison with experimental data are shown in  
section IV.  Finally,  the summary and the outlook are presented in section V.

\section{Space-time evolution of QGP}
The QGP  formed in $pp$ and AA collisions with high internal pressure
will expand in space with time. The  evolution 
of QGP is governed by the hydrodynamics. Here we use 
the second order relativistic viscous hydrodynamics.
For solving the hydrodynamical equations we need to specify the initial
conditions and the equation of state (EoS).  
In absence of any method based on the first principle to
estimate the initial temperature ($T_0$),  the following relation~\cite{rchwa}  has been used  in the present work to constrain $T_0$ by data:
\begin{equation}
T_{0} = \left[\frac{90}{g_{k} 4\pi^{2}}C^{\prime}\frac{1}{A_T\tau_0}1.5\frac{dN_{ch}}{dy}\right]^{1/3}
\label{t0}
\end{equation}
Although this equation is derived  by assuming isentropic expansion, we have 
used it also for the present case. We feel that the uncertainties
involved in this assumption at least for small viscosity, may not be large 
as compared to the uncertainties
involved in the other quantities like transverse area of the
system ($A_T$), statistical degeneracy ($g_k$) of the QGP phase and
the thermalization time ($\tau_0$).   
In Eq.~\ref{t0}, $C^{\prime} = \frac{2\pi^{4}}{45\zeta(3)} \approx 3.6$.  
We have also taken, $\frac{dN_{ch}}{dy} \cong \frac{dN_{ch}}{d\eta}$
which is exact in the massless limit.

\begin{figure}[ht!]
\includegraphics[scale=0.354]{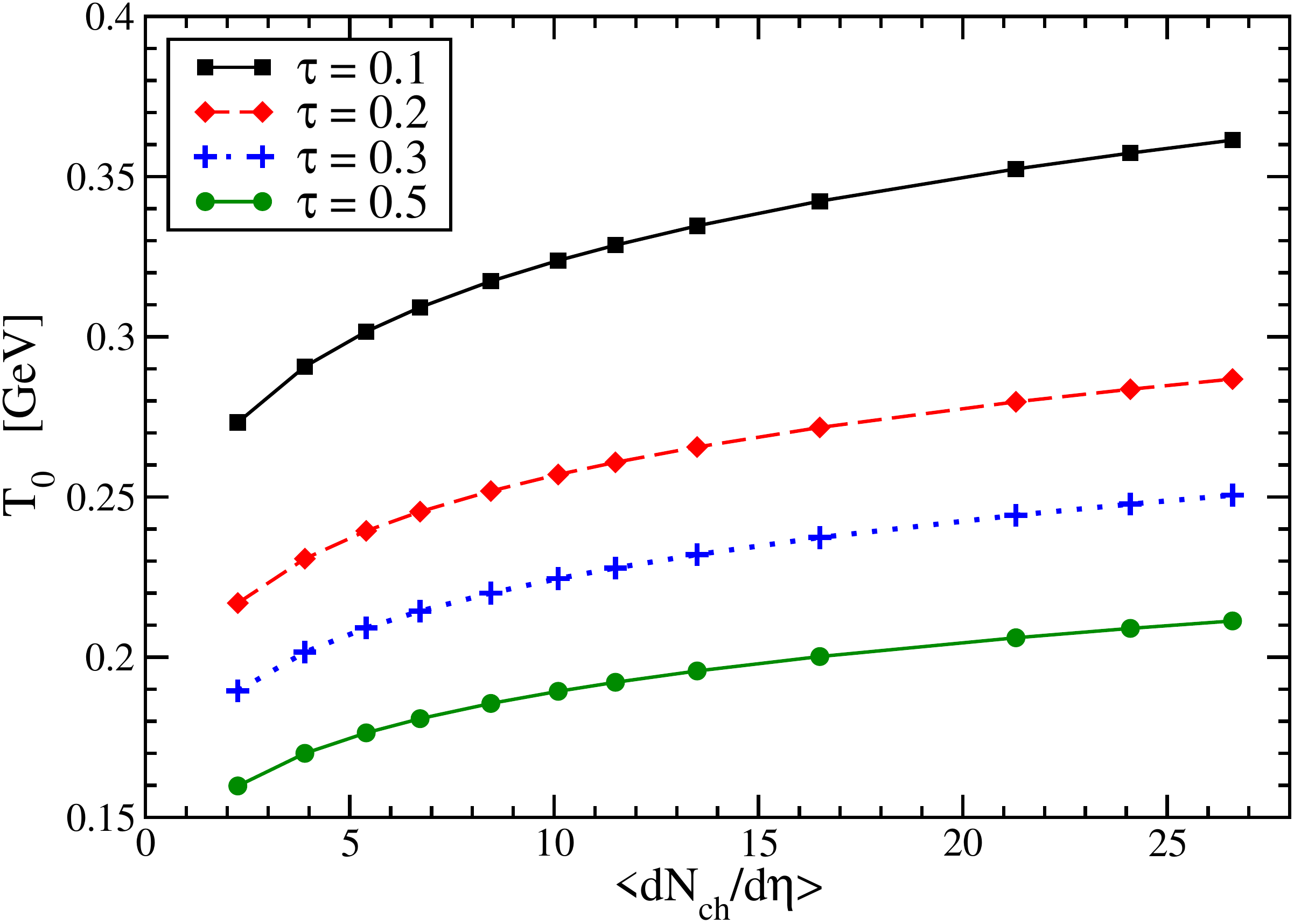}
\caption{The variation of the initial temperature as a function of final state charged-particle multiplicity.
}
\label{temp0}
\end{figure}

Here the transverse overlap area $A_{T} = \pi R_{T}^{2}$, is computed by using IP-Glasma  model, 
where $R_{T}$ is the transverse radius of the fireball~\cite{at}. This description of the transverse area is 
based on the impact parameter defined for $pp$ collisions and combined with a description of the particle 
production based on the color glass condensate model. It is worthwhile to mention that initial 
thermalization  time ($\tau_{0}$) plays a key role in estimating the $T_0$.
While, there is no way to estimate the $\tau_{0}$ from first principle but it may be  reasonable 
to assume that thermalization time will decrease with increasing center-of-mass collision energy,
i.e. $\tau_{0}\propto 1/\sqrt{s}$ ~\cite{hwa2,hwa3}. Here  we consider $\tau_{0} =  0.3,\; 0.2\; \&\; 0.1$ fm 
respectively for $\sqrt{s} = 5.02,\;7.0,\; \& \; 13$ TeV  for $pp$ collisions. 
Corresponding to these  values of $\tau_0$, we obtain the values of $T_0$ as a function of
multiplicity. The results have been depicted in Fig.~\ref{temp0}. \\

Fig.~\ref{temp0}, emphasizes the impact of the thermalization time ($\tau_{0}$) on the initial 
temperature ($T_{0}$). For $\tau_{0} = 0.1$ to $0.3$ fm, we get $T_{0}\sim 200$ to $275$ MeV at 
low multiplicities. At high multiplicity $pp$ collision at $\sqrt{s} = 13$ TeV, initial temperature 
reaches a value $\sim 370$ MeV which is comparable to the initial temperature achieved
in heavy-ion collisions. However, $T_0$ obtained from Eq.~\ref{t0}, is a slowly varying function 
of the multiplicity ($T_0\sim (dN/dy)^{1/3}$). 
Therefore, we get $T_{0} \simeq 200$ MeV at low multiplicity in $pp$ collision 
at $\sqrt{s} = 5.02$ TeV. However,  we find that $T_{0}>T_{c}$,  
suggesting  that QGP-like medium may be produced even at low multiplicities in $pp$ collisions at LHC energies, $T_c \approx 155$ MeV is predicted in the Lattice QCD calculation~\cite{bazavov}.  It is expected that in such a hot medium formed at $\sqrt{s}=13$ TeV, say,
the $J/\psi$ will be suppressed.

\subsection{Variation of temperature ($T$) with proper time ($\tau$)}
The temperature of the system produced in ultra-relativistic 
heavy-ion collisions decrease with time. 
The rate of the decrease of  temperature 
can be obtained  by solving the  
second-order hydrodynamical 
equation~\cite{amuronga}. These second order equations are derived 
from kinetic theory using Grad's 14-moment approximation method 
within the framework of 
the M\"{u}ller-Israel-Stewart equation~\cite{gam,mis1,mis2,mis3}
which in (1+1) dimension with boost invariance reads:

\begin{equation}
\frac{dT}{d\tau} = -\frac{T}{3\tau} + \frac{T^{-3}\phi}{12a\tau}
\label{so1}
\end{equation}
and 
\begin{equation}
\frac{d\phi}{d\tau} = -\frac{2aT\phi}{3b} - \frac{1}{2}\phi\left[\frac{1}{\tau} -\frac{5}{\tau}\frac{dT}{d\tau}\right] + \frac{8aT^{4}}{9\tau}
\label{so2}
\end{equation}

The $\phi$ in Eq.~\ref{so2} measures the change in the shear 
viscosity ($\eta$) with time $\tau$. In short, $\phi$ indicates the  
characteristics of the medium formed under extreme conditions in ultra-relativistic 
collisions. For the first order solution of the Eq.\ref{so1}, the $\phi$ is defined as, 
$\phi = 4\eta/3\tau$. The constants $a$ and $b$ are given by,

\begin{equation}
a = \frac{\pi^{2}}{90} \left[ 16 + \frac{21}{2}N_{f} \right]
\label{a}
\end{equation}
and
\begin{equation}
b = (1 + 1.70 N_{f})\frac{0.342}{(1 + N_{f}/6)\alpha_{s}^{2}\ln(\alpha_{s}^{-1})}
\label{b}
\end{equation}
where $N_{f} = 3$, is the number of flavors and $\alpha_s$ is the strong coupling constant.
We have taken $\alpha_s=0.5$ here as in Ref.~\cite{amuronga}. We have used
an EoS with conformal symmetry ($P=\epsilon/3$).  The bulk
viscosity is zero for such EoS.  

For an ideal system, the proper-time dependence of temperature is given by,
\begin{equation}
T(\tau,b) = T_{0}\left(\frac{\tau_{0}}{\tau}\right)^{1/3}
\label{it}
\end{equation}

However, for a viscous system, the Eqs.~\ref{so1} \& ~\ref{so2} have been solved numerically.
We have obtained the initial condition for $\phi$ with the help of quasi-particle model equation of state (QPM EoS) where
QGP is considered as viscous medium.
In  QPM EoS, entropy density, $s = c + d T^{3}$ here $c = 0.00482\;\; $ GeV$^{3}$ and $d = 16.46$ 
are the fit parameters~\cite{capt3}. By taking the lower bound of $\eta/s=1/(4\pi)$ from  Ads/CFT, the value of $\phi$ at $\tau_{0}$ is
obtained as: $\phi_{0} = \frac{1}{3\pi}\frac{s_{0}}{\tau_{0}}$, here $s_{0} = c + d T_{0}^{3}$.
The variation of $T$ with $\tau$ is crucial to determine the 
lifetime of QGP and hence for the signal of QGP. This variation is 
sensitive to $\phi_0$ as demonstrated in Ref.~\cite{MurongaPRC}.  
We have taken the low value of $\eta/s$ to set the intial condition
for $\phi(\tau_0)=\phi_0$. The lower $\eta/s$ will give faster cooling
and hence a smaller lifetime of the thermal QGP system. 
Therefore, if the suppression is observed for such a conservative 
scenario (QGP with a small lifetime), then the realization of
thermal system in pp collisions will be better assured.
\\

In order to predict the transverse expansion effect on the medium evolution, transverse expansion is incorporated as a correction to the
(1$+$1) dimension cooling rate. It is assumed that transverse expansion starts at a time ($\tau_{tr}$) as: $\tau_{tr} = \tau +  0.2928 (r/{c_{s}})$~\cite{capt3,naka,lokh}. Here $r$ is the transverse distance and $c_s$ is the speed of sound in QCD medium. Expansion rate in transverse direction is estimated by replacing $\tau_{tr}$ in coupled cooling rate Eqs.~\ref{so1} \& ~\ref{so2}.

\begin{figure}[h!]
\includegraphics[scale=0.354]{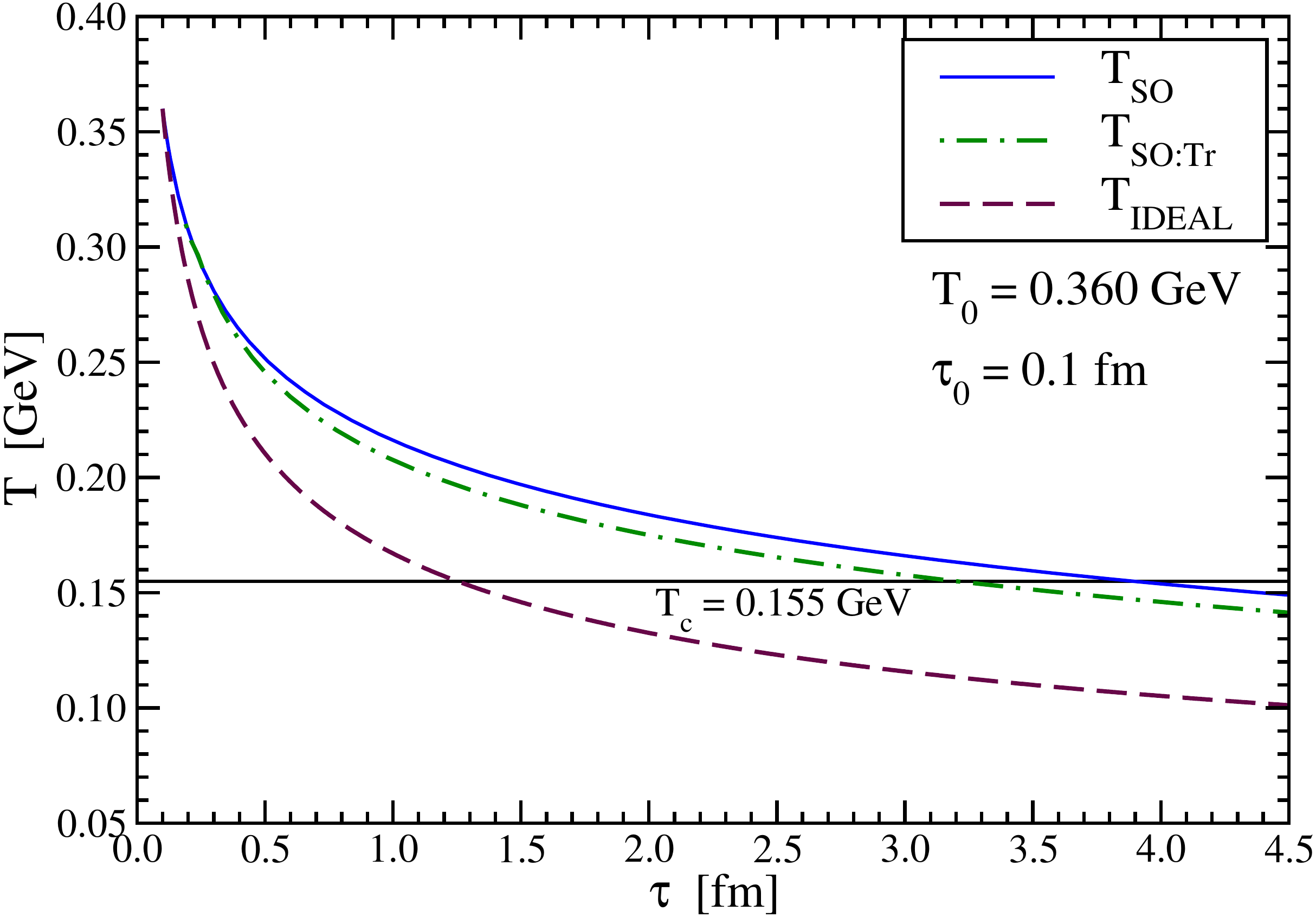}
\caption{Variation of temperature with proper time ($\tau$) is 
shown here for ideal (dashed line), second order viscous hydrodynamics (solid line) and second order viscous hydrodynamics with transverse expansion (green dashed line). 
}
\label{temp}
\end{figure}

In Fig.~\ref{temp}, the variation of temperature with
proper-time is displayed for viscous 
QGP (solid blue line), viscous QGP 
with the inclusion of transverse expansion (green dashed line) and ideal 
QGP (violet dashed line) for system produced in $pp$ collisions at $\sqrt{s}=13$ TeV. 
The generation of heat due to dissipation makes the cooling of viscous QGP slower. However, 
considering the transverse expansion correction to current formulation, marginally 
increases cooling 
rate compared to (1+1)-dimension expansion
as the lifetime of the QGP in smaller system 
is shorter than bigger system produced in nuclear collisions. 
Transverse temperature cooling rate, 
$T_{SO:Tr}$, mentioned in Fig.~\ref{temp}, uses initial transverse position $r = 0.2$ fm (say) at $\tau = \tau_{0}$. Change in the viscous cooling rate after including transverse correction is not much prominent. While cooling for ideal QGP medium is large  throughout medium evolution as compare to both (with and without transverse correction) viscous cases. Solution of the second-order hydrodynamics excluding transverse expansion shows that the QGP lifetime will be enhanced (3.8 fm/c) compared to the  ideal fluid (1.5 fm/c). Such an increase in lifetime will be crucially important for the signal of QGP.

\subsection{Effective Temperature}
Charmonia being a massive object, does not become a part of the thermalized medium which  induces a temperature gradient between the medium and charmonium. 
This can be measured by using the relativistic Doppler shift effect, which comes into the existence due to the velocity difference between medium particles and 
the charmonium. 
Therefore, relative velocity ($v_{r}$) between medium and charmonia is calculated to obtain an effective temperature felt by the charmonium. 
The velocities of the medium and charmonium are denoted by $v_{m}$ and $v_{J/\psi(nl)}$, respectively. This relativistic Doppler shift causes an 
angle dependent effective temperature ($T_{eff}$) expressed as
(see Refs.~\cite{teff, wols2} for details):

\begin{equation}
T_{eff}(\theta,|v_{r}|) = \frac{T(\tau,b)\;\sqrt{1 - |v_{r}|^{2}}}{1 -
|v_{r}|\;\cos \theta}, 
\label{tt}
\end{equation}
where $\theta$ is the angle between $v_{r}$ and incoming light partons. 
To calculate the  $v_{r}$, we have taken medium velocity, $v_{m} \sim 0.7c$, and charmonium velocity 
$v_{J/\psi(nl)} = p_{T}/E_{T}$, where $p_T$ is transverse momentum of charmonia and $E_T=\sqrt{p_T^2 + M_{nl}^2}$ 
is its transverse energy, $M_{nl}$ is the mass of the charmonium state. 
We have taken the average of Eq.(\ref{tt}) over the solid angle and 
obtained the average effective temperature as given by:
\begin{equation} 
T_{eff}(\tau,b,p_{T}) = T(\tau,b)\;\frac{\sqrt{1 -
|v_{r}|^{2}}}{2\;|v_{r}|}\;\ln\Bigg[\;\frac{1 + |v_{r}|}{1 - |v_{r}|}\Bigg]\,.
\end{equation}

\section{Modification in  $J/\psi$ and $\psi(2S)$ yield in the QGP Medium}
The formulation of the current work is based on a model,  UMQS recently proposed in~\cite{capt3}. The UMQS is 
described below briefly for the sake of completeness with particular emphasize on the 
modifications wherever required. 

\subsection{Charmonium Kinematics in Evolving QGP}

The charmonia during their propagation through evolving medium  get influenced by the medium in several ways. 
The kinematics of the charmonium ($J/\psi(nl)$) and  $c\bar c$ 
pairs formation and recombination with the evolution of the QGP medium, can be written by using the transport equation as:
\begin{equation}
 \frac{d N_{J/\psi(nl)}}{d\tau} = \Gamma_{F,nl} N_{c}~N_{\bar{c}}~[V(\tau)]^{-1} - \Gamma_{D,nl} N_{J/\psi(nl)}.
\label{tq}
\end{equation}

The first term in 
the right hand side of
Eq.(\ref{tq}), is a formation term and second one corresponds to the dissociation. 
$\Gamma_{F,nl}$ and $\Gamma_{D,nl}$ are the recombination  and dissociation rates corresponding to the  
regeneration and dissociation of $J/\psi(nl)$, respectively. $V(\tau)$ is the volume of the medium.
We assume that initially,
the number of charm ($N_{c}$) and anti-charm quarks $(N_{\bar{c}})$ are produced in equal numbers, 
$N_{c}$ = $N_{\bar{c}}$ = $N_{c\bar{c}}$. The Eq.(\ref{tq}) can be  solved analytically under the 
assumption of $N_{J/\psi}(nl) < N_{c\bar{c}}$ at $\tau_{0}$~\cite{thews1,thews2}: 
\begin{align}
N_{J/\psi(nl)}(\tau_{QGP},b,p_{T})\; = \; \epsilon_1(\tau_{QGP},b,p_{T}) \bigg[
N_{J/\psi(nl)}(\tau_{0},b)\;\; \nonumber
\\ + N_{c\bar{c}}^{2} \int_{\tau_{0}}^{\tau_{QGP}} \Gamma_{F,nl}(\tau,b,p_{T})
[V(\tau,b)\epsilon_2(\tau,b,p_{T})]^{-1} d\tau \bigg]
\label{tq1}
\end{align}
Here $N_{J/\psi(nl)} (\tau_{QGP},b,p_{T})$ is the net number of charmonium formed during the
evolution of QGP, $\tau_{QGP}$. 
The $N_{J/\psi(nl)}(\tau_{0},b)$ is the number of initially produced charmonium at various 
centralities~\footnote{The impact parameter ($b$)-dependent centrality bins for $pp$ collisions are obtained by using the formulation of Ref.~\cite{suman}.} at time $\tau_0$. 
We have calculated the $N_{J/\psi(nl)}(\tau_{0},b)$ corresponding to the centrality bins:
\begin{equation}
N_{J/\psi(nl)}(\tau_{0},b) = \sigma_{J/\psi(nl)}^{NN}\; T_{pp}(b),
\end{equation}
where, $T_{pp}(b)$ is the overlap function 
for $pp$  obtained by using the Glauber model as demonstrated in~\cite{suman}. 
We have obtained the number of charm and anti-charm quarks 
given by, $N_{c\bar{c}} = \sigma_{c\bar{c}}^{NN}\; T_{pp}(b)$. 
The values of $\sigma_{J/\psi(nl)}^{NN}$ and $ \sigma_{c\bar{c}}^{NN}$, used in the calculation,
are given in Table~\ref{tbI}:
\begin{table}[ht!]
\caption{The values of $\sigma_{J/\psi(nl)}^{NN}$ and $ \sigma_{c\bar{c}}^{NN}$ cross-sections at mid-rapidity~\cite{ts,ts2}.}
\vspace{2mm}
\begin{ruledtabular}
\begin{tabular}{ccccccc}
$\sqrt{s}$ (TeV)&$\sigma_{J/\psi}^{NN}(\mu b)$&$\sigma_{\chi_{c}}^{NN}(\mu b)$&$\sigma_{\psi}^{NN}(\mu b)$&$\sigma_{c\bar{c}}^{NN}(mb)$\\
\hline\\
pp@5.02& $10.152$ & $3.045$ & $1.015$ & $ 2.097$\\
\\
pp@7& $12.42$ & $3.726$ & $1.242$ & $ 2.424$\\
\\
pp@13& $16.72$ & $5.016$ & $1.672$ & $ 2.942$\\
\end{tabular}
\end{ruledtabular}
\label{tbI}
\end{table}

In Eq.(\ref{tq1}), $\epsilon_1(\tau_{QGP})$ and $\epsilon_2(\tau)$ 
are the decay  factors for the meson due to 
gluonic dissociation and collisional damping in QGP 
(of lifetime $\tau_{QGP}$) 
and $\tau$ is the evolution time, respectively. 
These factors are obtained by using the following expressions:
\begin{equation}
\epsilon_1(\tau_{QGP},b,p_{T})= \exp{\left[-\int_{\tau_{nl}^{'}}^{{\tau_{QGP}}}
\Gamma_{D,nl} (\tau,b,p_{T}) \;d
\tau\right]},
\end{equation}
and 
\begin{equation}
\epsilon_2(\tau,b,p_{T}) =
\exp{\left[-\int_{\tau_{nl}^{'}}^{{\tau}}\Gamma_{D,nl}(\tau^{\prime},b,p_{T}) \; d
\tau^{\prime}\right]}.
\end{equation}

Here, $\Gamma_{D,nl}(\tau,b,p_{T})$ is the sum of collisional damping
and gluonic dissociation decay rates, discussed in the following sections.
The lower limit of the above integral ($\tau_{nl}^{'}= \gamma\tau_{nl}$, here $\gamma$ is Lorentz factor) 
is taken as the charmonium dilated formation time where the 
dissociation due to color screening becomes negligible. In the equilibrated scenario of the QGP, 
these dissociation factors strongly depends on the evolution of the medium.

\subsection{Suppression Mechanisms}

The charmonium floating in medium gets dissociated into its 
constitutions due to QGP medium effects and as a result the initial production 
of charmonia gets suppressed. 
The input parameters used in the model for calculating the charmonium suppression in the QGP medium  
are given in the Table~\ref{tbII}.\\

\begin{table}[h!]
\caption{The values of mass ($M_{nl}$), dissociation temperature ($T_{D}$) and formation time ($\tau_f$) are taken from Refs.~\cite{nendzig,ganesh}.}
\vspace{2mm}
\begin{ruledtabular}
\begin{tabular}{ccccccc}
	      &$J/\psi$&$\chi_{c}$&$\psi(2S)$  \\
\hline\\
$M_{nl}$ (GeV)& 3.1 & 3.5 & 3.7\\
\\
$T_{D}$ (MeV) & 325  & 180 & 173\\
\\
$\tau_{f}$ (fm)&0.89 & 2.0 & 1.5 \\
\end{tabular}
\end{ruledtabular}
\label{tbII}
\end{table}

In the following sections we describe the suppression mechanisms in brief along with the 
regeneration process which causes reduction in the  suppression.

\subsection{Color Screening}

The suppression mechanism due to color screening
was first proposed by Matsui and Satz~\cite{matsui}.
They proposed that like electric charge screening in QED plasma, the partons in the 
QGP medium screens the color charges.  
The screening  
prevents the formation of the
$c\bar{c}$ bound states~\cite{chu}. The color screening of the real part of 
the quark-antiquark potential is an independent suppression mechanism that dominates 
in the initial phase of QGP where the medium temperature is very high. Original color
screening mechanisms have gone through many refinements. In this work, the 
pressure  is parameterized in the transverse plane instead of 
energy density~\cite{mishra,pks1,pks2}. In this model it is  assumed that pressure 
vanishes at the phase boundary, i.e. at $r = R_{T}$, where $R_{T}$ is the transverse radius of cylindrical QGP. 

\begin{equation}
 p(\tau_{0},r) = p(\tau_{0},0)h(r)
\end{equation}
where,
\begin{equation}
 h(r) = \left(1 - \frac{r^{2}}{R_{T}^{2}}\right)^{\beta} \theta (R_{T} - r)
\end{equation}

The factor $p(\tau_{0},0)$ is obtained in the refs.~\cite{pks1,pks2}. $h(r)$ is the radial distribution function in transverse 
direction and $\theta$ is the unit step function. The exponent $\beta$ in the above equation depends on the energy deposition 
mechanism whose dependence on pressure is shown in Ref.~\cite{capt3}.

The variation of pressure as the function of $\tau$ is given by:\\
\begin{equation}
p(\tau,r) = A + \frac{B}{\tau^q} + \frac{C}{\tau} + \frac{D}{\tau^{c_s^2}}
\end{equation}
where A = -$c_1$, B = $c_2c_s^2$, C = $\frac{4\eta q}{3(c_s^2 - 1)}$ and $D = c_3$, here  $c_1$, $c_2$, $c_3$ are constants and have been calculated
using different boundary conditions on energy density and pressure. Other parameters appeared above are: $c_{s}$ is speed of sound in QGP medium, 
$\eta$ is shear viscosity of the medium and $q = c^{2}_{s} + 1$.
Determining the pressure profile at initial time $\tau = \tau_{0}$ and at screening time $\tau = \tau_{s}$, we get: 
\begin{equation}
p(\tau_{0},r) = A + \frac{B}{\tau_{0}^q} + \frac{C}{\tau_{0}}
+ \frac{D}{\tau_{0}^{c_s^2}} = p(\tau_{0},0)\,h(r)
\end{equation}
\begin{equation}
p(\tau_s,r) = A + \frac{B}{\tau_s^q} + \frac{C}{\tau_s} +
\frac{D}{\tau_s^{c_s^2}} = p_{QGP}
\end{equation}
where $p_{QGP}$ is the pressure of the QGP phase.
Putting variation of $T$ and $P$ with $\tau$ we have determined the radius of the screening region, 
$r_{s}$. The screening radius defines a region where effective medium 
temperature of a particle is larger than the 
dissociation temperature ({\it i.e.} $T_{eff} \ge T_{D}$). If $T_{eff} < T_{D}$, then $r_{s} \rightarrow 0$ 
which suggests that melting of the quarkonia due to color screening would be negligible in such a 
situation.\\

The $c\bar c$ pairs formed inside the screening region at a point $\vec
r_{J/\psi}$, may escape the region, if $|\vec{r_{J/\psi}}\; +\;
\vec{v_{J/\psi}}t_{f} |\; > \;r_{s}$. Here $v_{J/\psi} = p_{T} / E_{T}$, is
charmonium velocity. The condition for escape of $c\bar c$ pair is expressed as:

\begin{equation}
\cos(\phi) \geq Y; \; \; Y = \frac{(r_{s}^{2} - r_{J/\psi}^{2})M_{nl}-
\tau_{nl}^{2} p_{T}^{2} / M_{nl}}{2\;r_{J/\psi}\; p_{T}\;
\tau_{nl}},
\label{phis}
\end{equation}
where, $\phi$ is the angle between the velocity ($\vec{v_{J/\psi}}$) 
and position vector ($\vec{r_{J/\psi}}$), and $m$ is mass of a particular charmonium
state.\\ 

Based on Eq.(~\ref{phis}), the allowed values of the azimuthal angle, $\phi_{max}(r)$ for survival of charmonium is expressed as:

\begin{center}
 $\phi_{max}(r) =  \left\{ \begin{tabular}{c}
\vspace{2mm}
$\pi$  $\;\;$, if $\;\;$   $Y\leq -1$\\
\vspace{2mm}
$\pi - \cos^{-1}|Y|$ $\;\;$, if $\;\;$  $0\geq Y \geq -1$\\
\vspace{2mm}
$\cos^{-1}|Y|$ $\;\;$, if $\;\;$ $0\leq Y \leq -1$\\
\vspace{2mm}
 $0$ $\;\;$, if $\;\;$ $Y\geq 1$
\end{tabular}
\right\}$.
\end{center}
Here $r$ is the the radial separation between $\bar{c}$ and $c$ 
in the QGP medium.

The integration over $\phi_{max}$ along with radial separation $r$  gives the escape probability of $c\bar c$ pair from the screening region, 
which is defined as the survival probability.  The survival probability, $S_{c}^{J/\psi}$, for a  charmonium state is expressed as:

\begin{equation}
S_{c}^{J/\psi}(p_T,b)=\frac{2(\alpha_s+1)}{\pi R_T^2} \int_0^{R_T} dr\,
r\,\phi_{max}(r)\left\{1-\frac{r^2}{R_{T}^{2}} \right \}^{\alpha_s}
\label{cssp}
\end{equation}
where $\alpha_s=0.5$~\cite{chu,mishra}. 
The $R_{T}$ depends on the impact parameter $(b)$. 
We have calculated it by using the transverse 
overlap area $A_T$ as; $R_{T}(b) = \sqrt{A_{T}/\pi}$.
The value of $\alpha_s$ is chosen in such a way that beyond the chosen value, color screening mechanism becomes almost independent with respect to change in its values.

\subsection{Collisional Damping}
Collisional damping arises due to the inherent nature of the complex potential between ($c\bar c$) 
located  inside the QCD medium. The imaginary part of the potential in the 
limit of $t\rightarrow\infty$, represents the thermal decay width induced due to the  low frequency 
gauge fields that mediate interaction between two heavy quarks~\cite{laine}.\\

The charmonium dissociation due to collisional damping is obtained by using the effective 
potential models. In the model under  consideration, the singlet potential for $c\bar c$ bound state in the QGP 
medium is given by~\cite{nendzig,laine,aber}:

\begin{multline}
V(r,m_D) = \frac{\sigma}{m_D}(1 - e^{-m_D\,r}) - \alpha_{eff} \left ( m_D
+ \frac{e^{-m_D\,r}}{r} \right )\\
- i\alpha_{eff} T_{eff} \int_0^\infty
\frac{2\,z\,dz}{(1+z^2)^2} \left ( 1 - \frac{\sin(m_D\,r\,z)}{m_D\,r\,z} \right),
\label{pot}
\end{multline}

In Eq.~(\ref{pot}) the first and second term in the right hand side  are the string and the Coulombic terms, respectively. 
The third term is the
imaginary part of the heavy-quark potential responsible for the collisional damping. 
Here $\sigma$ is the string constant for $c\bar{c}$ bound
state, given as  $\sigma= 0.192$ GeV$^2$ and $m_{D}$ is 
the Debye mass, $m_D = T_{eff} \sqrt{4\pi \alpha_s^T \left(\frac{N_c}{3} + \frac{N_f}{6} \right) }$. 
Here $N_{c} = 3$, $N_{f} = 3$. The $\alpha_{s}^{T}$ is coupling
constant at hard scale, as it should be $\alpha_{s}^{T} = \alpha_{s}(2\pi T)\leq 0.50$. We have taken $\alpha_s^T \simeq 0.4430$ which is scaled at HTL energy 
$\sim 2\pi T$. 
The effective coupling constant, $\alpha_{eff}$ is defined 
at soft scale $\alpha_{s}^{s} = \alpha_{s} (m_{c} \alpha_{s}/2) \simeq 0.48$, given as  $\alpha_{eff} = \frac{4}{3}\alpha_{s}^{s}$.\\

The collisional damping, $\Gamma_{damp,nl}$ defines the charmonium decay induced due to 
the imaginary part of the complex potential. It is calculated using the first-order perturbation, 
by folding of imaginary part of the potential with the radial wave function:

\begin{equation}
\Gamma_{damp,nl}(\tau,p_{T},b) = \int[g_{nl}(r)^{\dagger} \left [ Im(V)\right] g_{nl}(r)]dr,
\label{cold}
\end{equation}
where $g_{nl}(r)$ is the charmonia singlet wave function. Corresponding to different
values of $n$ and $l$ (here $n$ and $l$ are the principal and the orbital quantum numbers), the wave functions 
can be obtained by solving the Schr\"{o}dinger equation for $J/\psi$, $\chi_{c}$, $\psi(2S)$.

\subsection{Gluonic Dissociation}
The dissociation of $J/\psi$ by gluons contributes to the 
reduced yields of $J/\psi$ in heavy-ion collisions.
The the cross-section for such dissociation process is given by~\cite{nendzig}:
\begin{multline}
\sigma_{d,nl}(E_g) = \frac{\pi^2\alpha_s^u E_g}{N_c^2}\sqrt{\frac{{m_{c}}}{E_g
+ E_{nl}}}\\
\;\;\;\;\;\times \left(\frac{l|J_{nl}^{q,l-1}|^2 +
(l+1)|J_{nl}^{q,l+1}|^2}{2l+1} \right),
\end{multline}
where $J_{nl}^{ql^{'}}$ is the probability 
density obtained by using the singlet
and octet wave functions as follows: 
\begin{equation}
 J_{nl}^{ql'} = \int_0^\infty dr\; r\; g^*_{nl}(r)\;h_{ql'}(r).
\end{equation}
where $m_c=1.5$ GeV, is the mass of the charm quark and 
$\alpha_s^u \simeq 0.59$, is the coupling constant, scaled as 
$\alpha_{s}^{u} = \alpha_{s}(\alpha_{s}m_{c}^{2}/2)$. The $E_{nl}$ is the energy eigen values 
corresponding to the charmonium wave function, $g_{nl}(r)$. The  octet wave function $h_{ql'}(r)$ has been 
obtained by solving the Schr\"{o}dinger equation with the octet potential $V_{8} = \alpha_{eff}/8r$. The 
value of $q$ is determined by using the conservation of energy, $q = \sqrt{m_{c}(E_{g}+E_{nl})}$.

The gluonic dissociation rate, $\Gamma_{gd,nl}$ is obtained, by taking the thermal average of the 
dissociation cross-section~\cite{capt3}, 
\begin{equation}
\Gamma_{gd,nl}(\tau,p_{T},b) = \frac{g_d}{4\pi^2} \int_{0}^{\infty}
\int_{0}^{\pi} \frac{dp_g\,d\theta\,\sin\theta\,p_g^2
\sigma_{d,nl}(E_g)}{e^{ \{\frac{\gamma E_{g}}{T_{eff}}(1 +
v_{J/\psi}\cos\theta)\}} - 1},
\label{glud}
\end{equation}
where $p_T$ is the transverse momentum of the charmonium and $g_d = 16$ is the statistical degeneracy of the gluons. 
Now summing the decay rates corresponding to the collisional damping and the gluonic dissociation, we have calculated the combined effect in terms of 
total decay width denoted by, $\Gamma_{D,nl}(\tau,p_{T},b)$:
\begin{equation}
 \Gamma_{D,nl} = \Gamma_{damp,nl} + \Gamma_{gd,nl}.
\label{GmD}
\end{equation}

\subsection{Regeneration Factor}
There are two  processes by which charmonia can be 
reproduced in the QGP medium. The first one 
is  through the uncorrelated $c\bar c$ pairs present in the medium. They can recombine within the QGP medium at a later 
stage~\cite{pbm, andronic,rapp1,rapp2,thews1,thews2,thews3} to form a
charmonium state. This regeneration process is found  to be significant for charmonia in 
heavy-ion collisions at LHC energies because $c\bar c$  
are produced abundantly  
in the hot QGP medium. As $pp$ is a small system, the production of  $c\bar c$ will be small compared to the  heavy-ion collisions, therefore, the probability of regeneration due to uncorrelated $c\bar c$ is not considered 
in the present work. The regeneration due to correlated  $c\bar c$ pairs is the regeneration
mechanism which is just the reverse of gluonic dissociation. In this process, $c\bar c$ pairs may undergo a transition from color octet state to color 
singlet state in the due course of time in QGP medium. To account for the regeneration via correlated $c\bar c$ pairs in our current UMQS model, 
we have considered the de-excitation of octet state to singlet state via a gluon emission.
We have calculated this de-excitation in terms of recombination cross-section $\sigma_{f,nl}$ 
for charmonium by using the detailed balance to the gluonic dissociation cross-section 
$\sigma_{d,nl}$~\cite{capt3}:
\begin{equation}
 \sigma_{f,nl} = \frac{48}{36}\sigma_{d,nl} \frac{(s-M_{nl}^{2})^{2}}{s(s-4\;m_{c}^{2})}.
\end{equation}
where $s$ is the Mandelstam variable, related with the center-of-mass energy of $c\bar c$ pair, given as; $s = (p_c
+ p_{\bar{c}})^2$, where $ p_c$ and $ p_{\bar{c}}$ are four momenta of  $c$ and $\bar{c}$, respectively.\\
 
Finally we have obtained the recombination factor, $\Gamma_{F,nl}=<\sigma_{f,nl}\;v_{rel}>_{p_{c}}$ by taking the thermal average of the product of 
recombination cross-section and relative velocity $v_{rel}$ between $c$ and $\bar{c}$ :
\begin{equation}
 \Gamma_{F,nl} =
\frac{\int_{p_{c,min}}^{p_{c,max}}\int_{p_{\bar{c},min}}^{p_{\bar{c},max}}
dp_{c}\; dp_{\bar{c}}\; p_{c}^{2}\;p_{\bar{c}}^{2}\;
f_{c}\;f_{\bar{c}}\;\sigma_{f,nl}\;v_{rel}
}{\int_{p_{c,min}}^{p_{c,max}}\int_{p_{\bar{c},min}}^{p_{\bar{c},max}}
dp_{c}\; dp_{\bar{c}}\; p_{c}^{2}\;p_{\bar{c}}^{2}\;
f_{c}\;f_{\bar{c}}},
\end{equation}
where, ${p_c}$ and ${p_{\bar{c}}}$ are the $3$-momentum of charm and anti-charm
quark, respectively. The $f_{c,\bar{c}}$ is the modified Fermi-Dirac
distribution function of charm, anti-charm quark and expressed as; $f_{c,\bar{c}} =
\lambda_{c,\bar{c}}/(e^{E_{c,\bar{c}}/T_{eff}} + 1)$. Here $E_{c,\bar{c}} =
\sqrt{p_{c,\bar{c}}^{2} + m_{c,\bar{c}}^{2}}$ is the energy of the quarks and $\lambda_{c,\bar{c}}$ is their respective
fugacity factors~\cite{dks}. We have calculated the relative velocity of
$c\bar c$ pair in medium as follows:

\begin{equation}
v_{rel} =
\sqrt{\frac{({\bf p_{c}^{\mu}\;
p_{\bar{c} \mu}})^{2}-m_{c}^{4}}{p_{c}^{2}\;p_{\bar{c}}^{2}
+ m_{c}^{2}(p_{c}^{2} + p_{\bar{c}}^{2}  + m_{c}^{2})}}.
\end{equation}\\

Since the gluonic dissociation increases with the increase in temperature, it leads to the production 
of significant number of $c\bar c$ octet states in central collision where temperature is 
found  to be around $300$ MeV. Such that the de-excitation of $c\bar c$ octet states to $J/\psi$ 
enhance the regeneration of $J/\psi$ in central collisions as compared with the peripheral collisions.


\subsection{Net Yield}
The survival probability of charmonium ($S_{gc}^{J/\psi}$)  
in $pp$ collisions due to gluonic dissociation along
with collisional damping is given by: 
\begin{equation}
 S_{gc}^{J/\psi}(p_{T}, b) =
\frac{N_{J/\psi(nl)}^{f}(p_{T},b)}{N_{J/\psi(nl)}(\tau_{0},b)}\,.
\label{sp2}
\end{equation}

It is assumed here that at $\tau=\tau_0$ the color screening is the most dominating mechanism 
and would not allow for the charmonium to be formed. However, as
QGP cools down, its effect on quarkonia suppression becomes insignificant 
at the later stage of the evolution. 
The color screening is incorporated in the model as an independent mechanism with the other suppression
mechanisms in QGP. 
The net yield in terms of survival probability,  $S_{P}(p_{T},b)$ is expressed as:
\begin{equation}
 S_{P}^{J/\psi}(p_{T},b) = S_{gc}^{J/\psi}(p_{T}, b)\;S_{c}^{J/\psi}(p_{T},b).
 \label{spf}
\end{equation}

The feed-down formulation of the excited states of the charmonia, $\chi_{c}$ and $\psi(2S)$ into $J/\psi$ has been adopted from the Refs.~\cite{capt,nendzig,capt3}. To obtain the feed-down, here we have taken the ratio between the  net numbers of the initial and final $J/\psi$. The net initial number is obtained considering the feed-down of the higher resonances into $J/\psi$ in the absence of the QGP medium. This is given as $N_{J/\psi}^{in} = \sum_{J\ge I} C_{IJ} N(J)$, where $C_{IJ}$ is the branching ratio of state $J$ to decay into the state $I$. The net final number of $J/\psi$ includes medium effects in terms of survival probability ($S_{P}(p_{T},b)$) along with feed-down: $N_{J/\psi}^{fi} = \sum_{J\ge I} C_{IJ} N(J) S_{P}(J)$. The net generalized survival probability including feed-down correction can be written as;

\begin{equation}
 S_{P}(I) = \frac{\sum_{J\ge I} C_{IJ} N(J) S_{P}(J)}{\sum_{J\ge I} C_{IJ} N(J)}
\end{equation}

It has been observed that only $60\%$ of $J/\psi$  originates from direct production whereas $30\%$ and $10\%$ from the decays of $\chi_{c}$ and $\psi(2S)$ respectively. The net survival probability of the $J/\psi$ after considering feed-down correction is given by,

 \begin{equation}
S_{P}^{f}=\frac{0.60\; N_{J/\psi}\; S_{P}^{J/\psi} + 0.30\; N_{\chi_{c}} S_{P}^{\chi_{c}} + 0.10\; N_{\psi}\; S_{P}^{\psi}}{0.60\;N_{J/\psi}+ 0.30\;N_{\chi_{c}} + 0.10\;N_{\psi}}. \end{equation}

\section{Results and Discussions}
The results obtained from the calculation mentioned above is contrasted with  the data available in the form of the $J/\psi$  yield corresponding to the normalized  multiplicities, defined as;
\begin{equation}
  N^{J/\psi}_{m} = \frac{dN_{J/\psi}/d\eta}{\langle dN_{J/\psi}/d\eta \rangle},\;\;\;\;\;\;\;\;\ N^{ch}_{m} = \frac{dN_{ch}/d\eta}{\langle dN_{ch}/d\eta \rangle}
\end{equation}
where $N^{J/\psi}_{m}$ and $N^{ch}_{m}$ are normalized by
the corresponding mean values in minimum bias $pp$ collisions.

\begin{figure*}[!ht]
\includegraphics[scale=0.445]{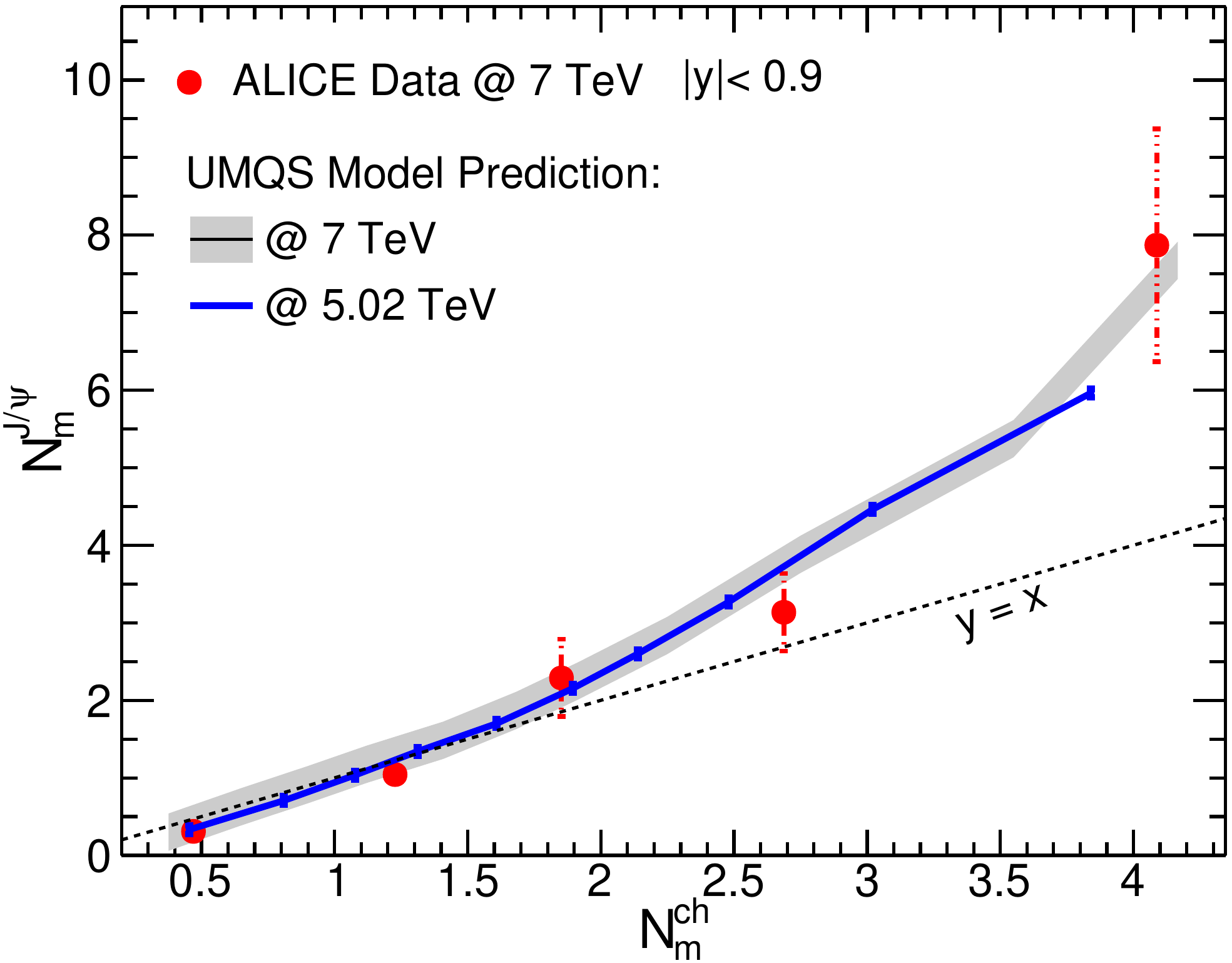}
\includegraphics[scale=0.445]{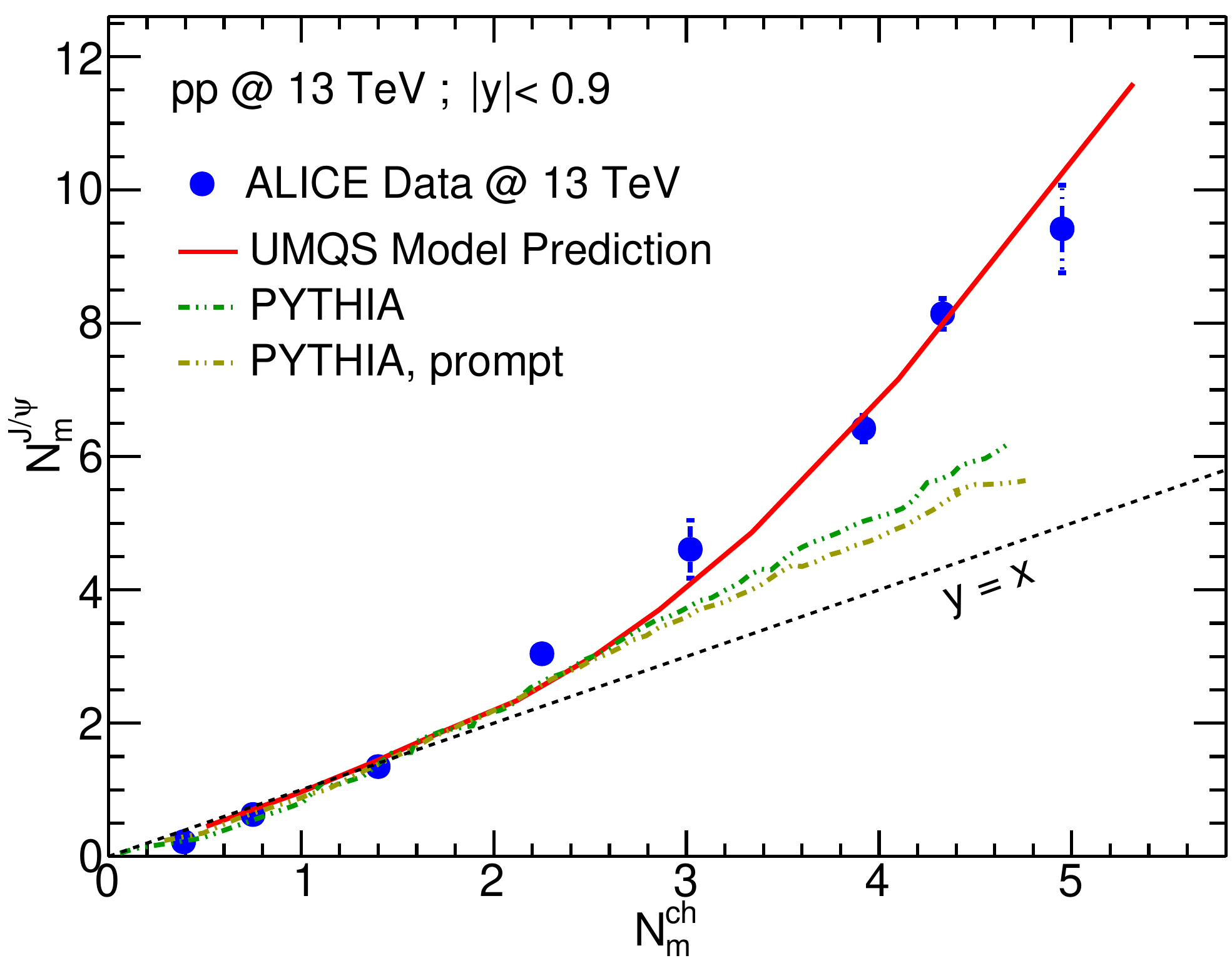}
\caption{Normalized inclusive $p_{T}$-integrated $J/\psi$ yield as a function of normalized charged-particle multiplicity  at mid-rapidity compared with ALICE data corresponding to V0M selection in $pp$ collisions at $\sqrt{s}= 7$ TeV~\cite{ALICE:2012pet}, prediction for $\sqrt{s}= 5.02$ TeV (left panel) and $\sqrt{s}= 13$ TeV~\cite{ALICE:2020msa} (right panel). PYTHIA 8.2 results from~\cite{ALICE:2020msa} are also included to contrast with the results of our model (right panel). The dashed line is a linear function with the slope of unity.} 
\label{fig:3}
\end{figure*}

The normalized $p_{T}$-integrated $J/\psi$ production 
as a function of normalized charged-particle multiplicity
is contrasted with the data from ALICE at mid-rapidity ($|y|< 0.9$) in $pp$ collisions
to check the feasibility of the UMQS model.
Left panel of Fig~\ref{fig:3} shows this comparison at $\sqrt{s} = 7$ TeV (data from~\cite{ALICE:2012pet}). 
The results from the present work for $\sqrt{s} = 5.02$ TeV is shown by blue line. 
Right panel of Fig~\ref{fig:3} shows the prediction of our model (red line) and 
PYTHIA 8.2 of ALICE data at  $\sqrt{s} = 13$ TeV~\cite{ALICE:2020msa}. 
PYTHIA 8.2 results are extracted from Ref.~\cite{ALICE:2020msa}. 
It is worth mentioning here that in 
Ref.~\cite{ALICE:2020msa}, the case of prompt $J/\psi$  production (predicted by PYTHIA 8.2)
is also included to illustrate the effect of non-prompt $J/\psi$ in the inclusive production. 
Clearly, both the results of PYTHIA 8.2 reproduce the experimental data reasonably well 
at low $N_{ch}$ but fails at higher $N_{ch}$ as shown in the right panel of Fig~\ref{fig:3}.\\

It is interesting to observe that UMQS  prediction explains the data within the uncertainties. From Fig~\ref{fig:3}, 
it is further observed that at high multiplicity the production of $J/\psi$ is larger than low multiplicity
regime. In UMQS  charmonium can be produced  through hard scattering and  recombination. 
The combined effect of both of these  production mechanisms dominates at high multiplicities. 
We have found that the regeneration plays an important role at high multiplicities while its contribution at low multiplicity bins is found 
to be negligible. The agreement of results displayed in Fig~\ref{fig:3} gives confidence to proceed with further analysis of quarkonia production 
by using UMQS in $pp$ collisions at the LHC energies.\\

Further, we have predicted the charmonium suppression in ultra-relativistic $pp$ collisions at the LHC energies. The UMQS model is 
used to calculate the
$p_{T}$ and centrality dependent survival probability of charmonium states formed
in $pp$ collisions at LHC energies at the mid-rapidity. 
The relative suppression of $\psi(2S)$ over $J/\psi$ is measured in the form of  double ratio or yield ratio, 
$S^{\psi(2S)}_{P}/S^{J/\psi}_{P}$ which is plotted as a function of multiplicity 
($dN_{ch}/d\eta$) 
and transverse momentum, ($p_{T}$).
It is to be noted that the ``S'', ``R'' and ``F''  stand for "suppression'', ``regeneration'' and  ``feed-down'', respectively. 
The results are shown with and without combining all the medium effects to demonstrate the significance of an individual mechanism.

\subsection{Centrality Dependent Suppression}
We have  obtained the centrality dependent survival probability for $J/\psi$ and $\psi(2S)$ by averaging over $p_{T}$
with the  $p_{T}$-distribution function $1/E_{T}^{4}$~\cite{mike}. 
The $p_{T}$-integrated centrality dependent survival probability is given by,

\begin{equation}
S_{P}(b) = \frac{\int_{p_{Tmin}}^{p_{Tmax}} d p_{T}  S_{P}(p_{T}, b)/(p_{T}^{2} + M_{nl}^{2})}{\int_{p_{Tmin}}^{p_{Tmax}} d p_{T}/(p_{T}^{2} + M_{nl}^{2})}
\end{equation}

The initial temperature obtained for low multiplicity exceeds the
value of the QCD transition temperature ($T_c$) obtained 
from lattice QCD calculation (see Fig.~\ref{temp0}). A marginal suppression of $J/\psi$ is obtained at low  multiplicity 
too and it increases with the final state multiplicities. However, $\psi(2S)$ is effectively more suppressed at low multiplicity as
compared with $J/\psi$. The effects of the individual 
mechanisms including feed-down are also shown in this section. 
The higher resonances of charmonium are more suppressed than $J/\psi$ and therefore the feed-down of these
states into $J/\psi$ enhances its suppression. However, regeneration mechanism reduces the suppression depending 
upon the medium size and temperature. For a relatively small QGP lifetime 
$\tau_{QGP}$ for $5.02$ TeV, the regeneration significantly reduces the suppression.\\

\begin{figure}[ht!]
\includegraphics[scale=0.354]{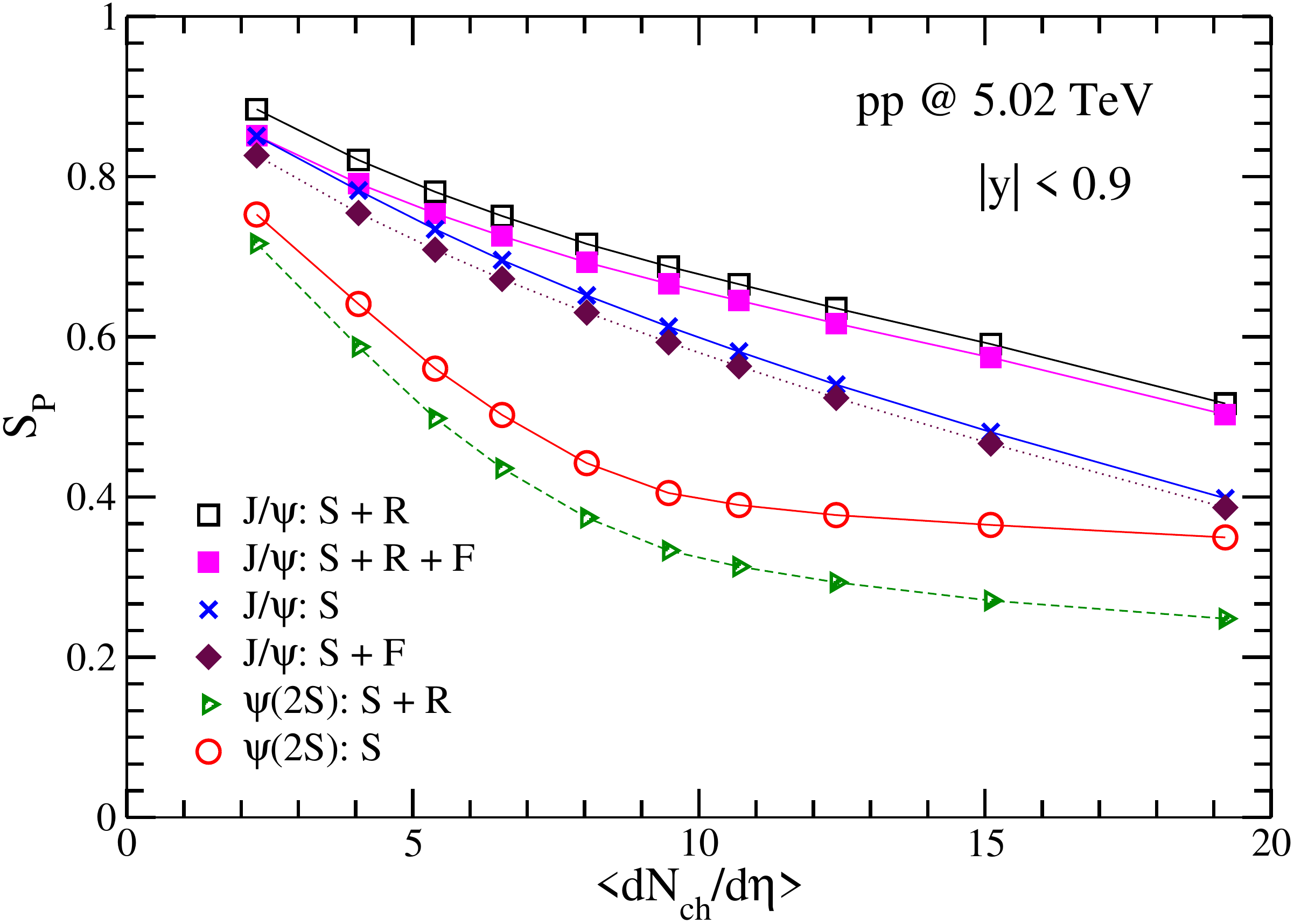}
\caption{The $p_T$-integrated survival probabilities of $J/\psi$ and $\psi(2S)$ are shown for $pp$ collisions at
$\sqrt{s}= 5.02$ TeV.}
\label{c211}
\end{figure}
In Fig.~\ref{c211}, charmonium suppression is shown as a function of the charged-particle multiplicity and the impacts of individual mechanisms
 are also presented. In this figure,
the effective regeneration is visible at high multiplicity while at low multiplicity, the effect of regeneration is found to be very small. Here, it is also shown 
that feed-correction enhances the $J/\psi$ suppression a bit. It is an indirect observation of the fact that higher resonances of charmonium are more suppressed than $J/\psi$. However, at 
$\sqrt{s} = 5.02$ TeV, we have observed an effective regeneration for $\psi(2S)$ which reduces its suppression significantly at high multiplicities. 
Inclusion of the 
regeneration along with feed-down correction  almost cancels out the feed-down effect and suggests the possibility of regeneration for the higher resonances at high multiplicities for $\sqrt{s} = 5.02$ TeV.\\

\begin{figure}[ht!]
\includegraphics[scale=0.354]{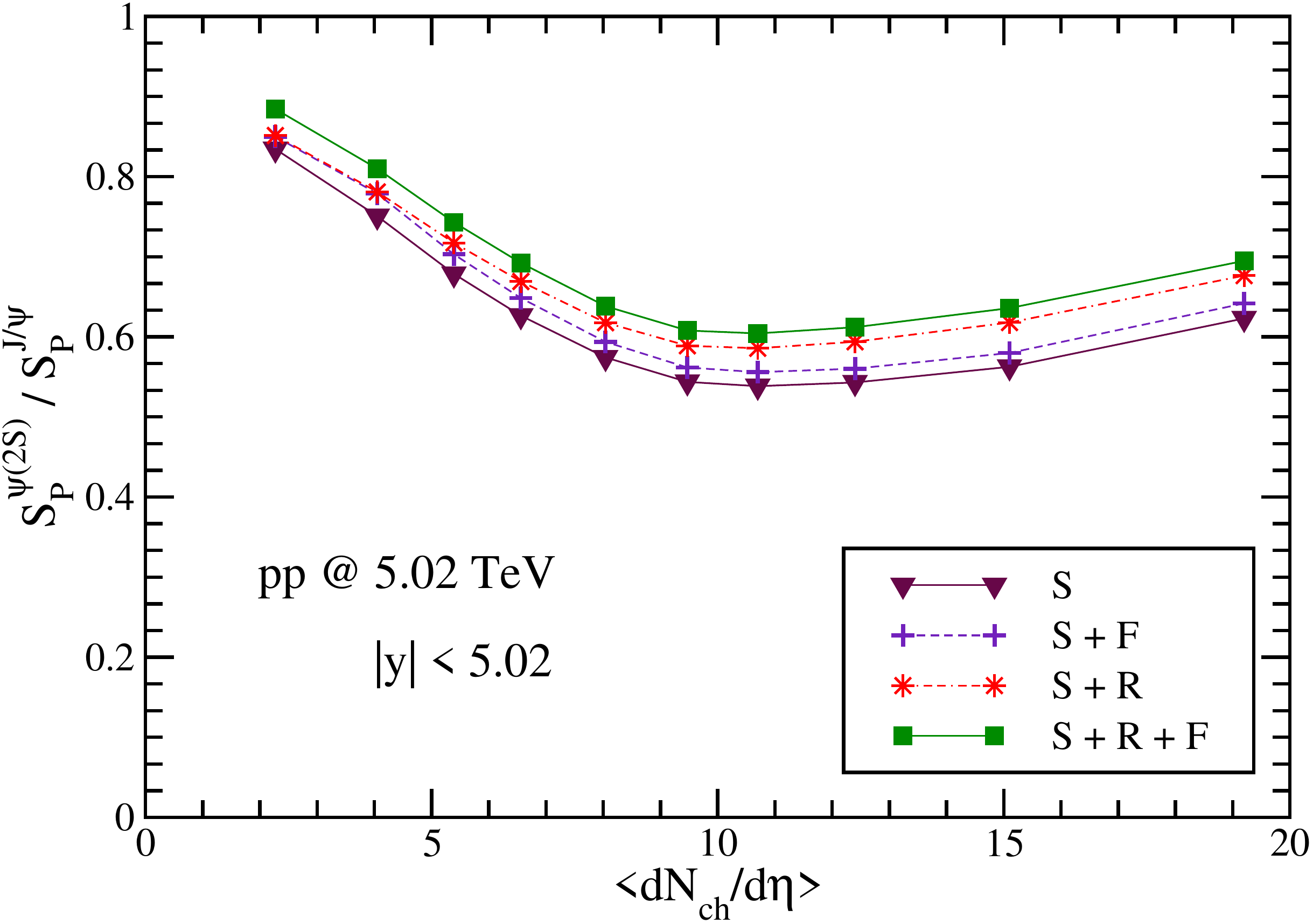}
\caption{The ratio of the survival probabilities of $\psi(2S)$ to $J/\psi$ as a function of final state event multiplicity
for $pp$ collisions at $\sqrt{s} = 5.02$ TeV.}
\label{c231}
\end{figure}

Fig.~\ref{c231} depicts the relative suppression of $\psi(2S)$ over $J/\psi$ including all the proposed medium effects. It shows, that at mid multiplicity bins $\psi(2S)$ is more suppressed as compared to $J/\psi$ for $\sqrt{s} = 5.02$. The coinciding point corresponding to ``S$+$F'' and ``S$+$R'' at the lowest multiplicity bin, suggests that the difference between $\psi(2S)$ and $J/\psi$ suppression is almost the same at $\sqrt{s} = 5.02$ TeV, if on one hand suppression for $J/\psi$ is increased by including feed-down correction and on other hand, suppression of $\psi(2S)$ is decreased by including regeneration.\\

\begin{figure}[ht!]
\includegraphics[scale=0.354]{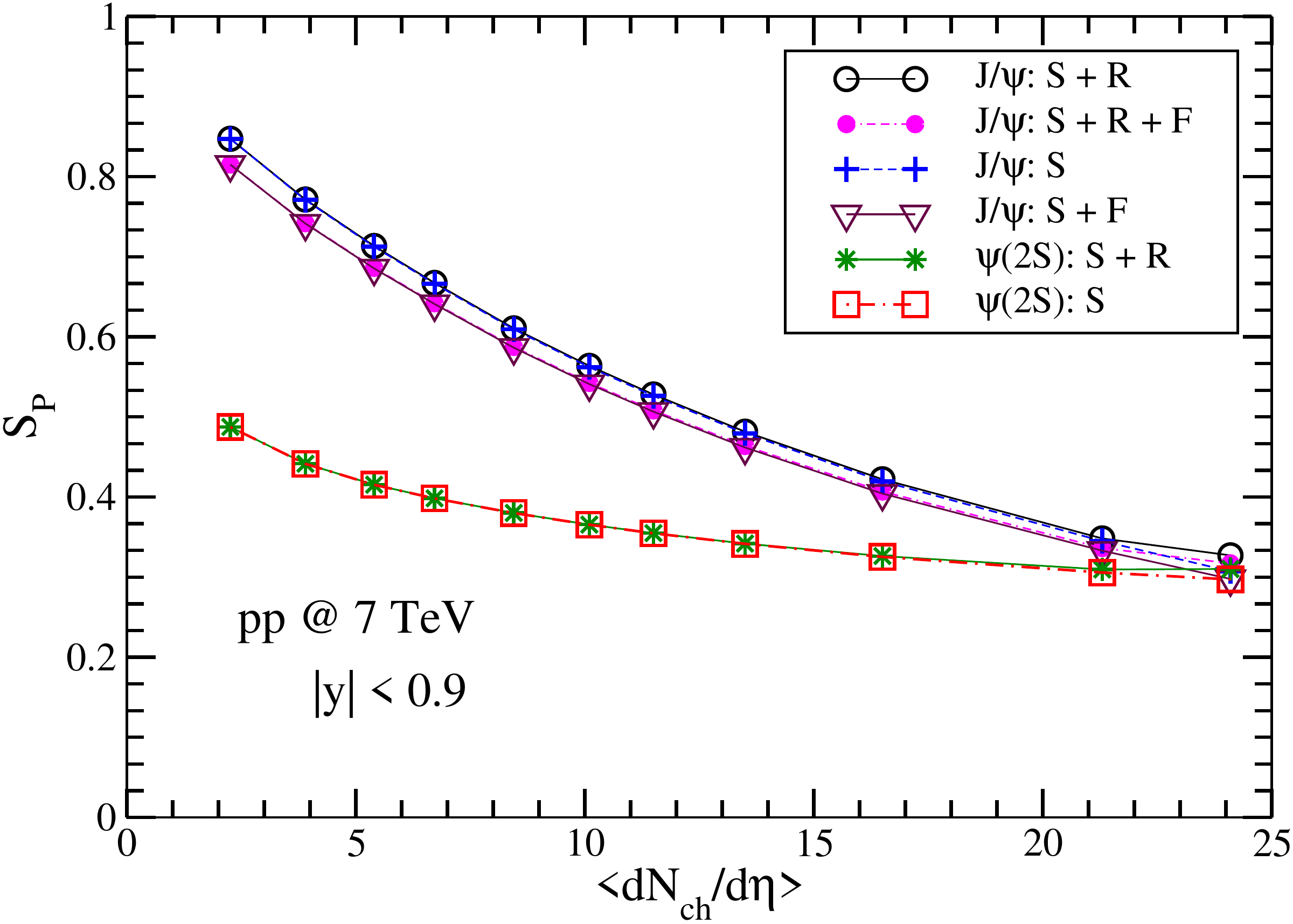}
\caption{The $p_T$-integrated survival probabilities of $J/\psi$ and $\psi(2S)$ are shown for $pp$ collisions at
$\sqrt{s}= 7$ TeV.}
\label{c212}
\end{figure}

\begin{figure}[ht!]
\includegraphics[scale=0.354]{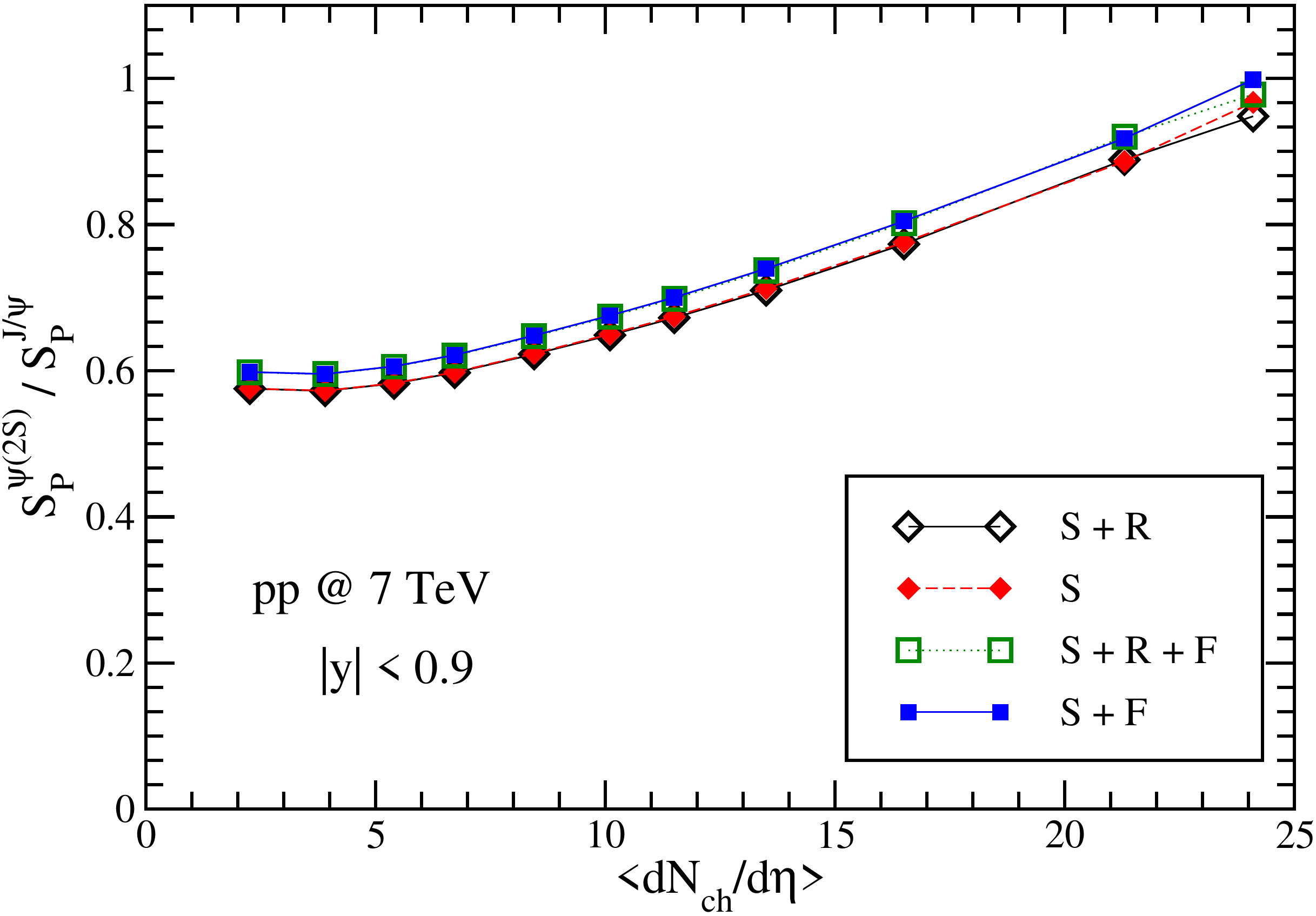}
\caption{The ratio of the survival probabilities of $\psi(2S)$ to $J/\psi$ as a function of final state event multiplicity
for $pp$ collisions at $\sqrt{s} = 7$ TeV.}
\label{c232}
\end{figure}

At $\sqrt{s} = 7$ TeV, $J/\psi$ and $\psi(2S)$ are almost equally suppressed at the high multiplicity, as shown in Fig.~\ref{c212}. 
However, $\psi(2S)$ is found to be more suppressed at low and mid multiplicities. Regeneration effect is also found to be negligible 
for $J/\psi$ and $\psi(2S)$ at $\sqrt{s} = 7$ TeV. An infinitesimal regeneration at very high multiplicity can be observed for 
$J/\psi$ and $\psi(2S)$ in Fig.~\ref{c212}. As regeneration being insignificant  at $\sqrt{s} = 7$ TeV, feed-down correction is 
found to take over the regeneration. Therefore, suppression along with regeneration and feed-down correction (``S+R+F'') 
follows the foot-step of the combined effects of the suppression and feed-down (``S+F''). Non-vanishing effect of regeneration at highest 
multiplicity bin, reduces the suppression a bit for $J/\psi$ and $\psi(2S)$ and almost takes away the effect of feed-down for $J/\psi$.\\

Similar result is displayed in Fig.~\ref{c232}, where at high multiplicity $\psi(2S)$ to $J/\psi$ 
yield ratio is approximately one, while at low multiplicities the ratio is less than one.
This indicates that $J/\psi$ is less suppressed than $\psi(2S)$ at peripheral $pp$ collisions, which correspond to lower final state event multiplicities. Results in Fig.~\ref{c232} also suggest a higher regeneration of  $J/\psi$ than $\psi(2S)$ at high multiplicity. 
The results for suppression and suppression along with feed-down, and regeneration are found to coincide at the high multiplicity domain.\\

\begin{figure}[ht!]
\includegraphics[scale=0.354]{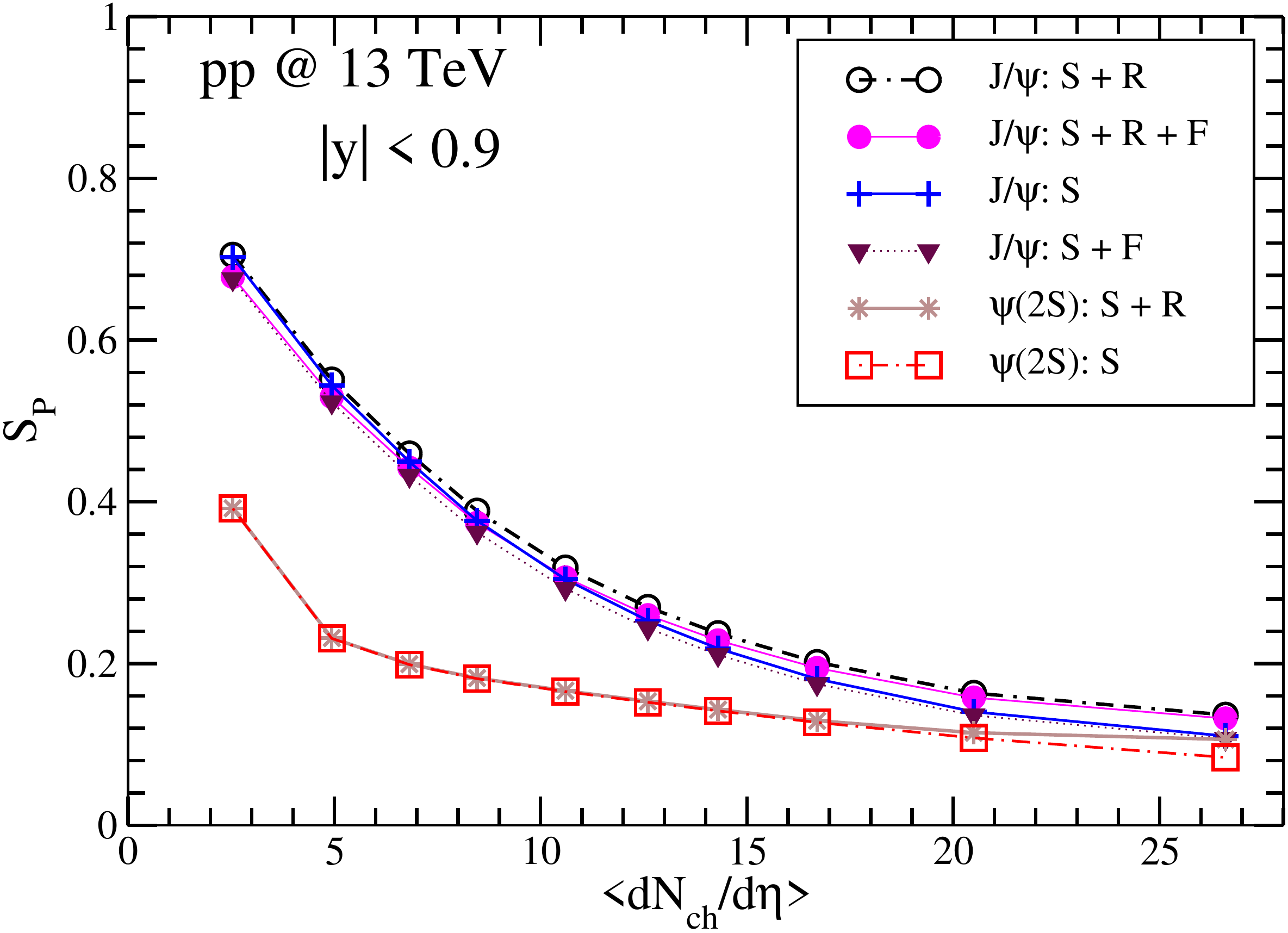}
\caption{The $p_T$-integrated survival probabilities of $J/\psi$ and $\psi(2S)$ are shown for $pp$ collisions at
$\sqrt{s}= 13$ TeV.}
\label{c213}
\end{figure}

It is expected that at $\sqrt{s}=13$ TeV the energy deposition is large 
which provides substantial  suppression of $J/\psi$ and $\psi(2S)$,
comparable to the heavy-ion collisions (see Fig.~\ref{c213}). 
Results displayed in Fig.~\ref{c213} show that the effect of regeneration is almost negligible at low multiplicity
(as in Fig.~\ref{c212}) while at high multiplicities it slightly reduces the suppression for both the charmonium states. 
$\psi(2S)$ is suppressed slightly more than $J/\psi$ at high multiplicity contrary to the results shown in Fig.~\ref{c212}. \\

\begin{figure}[ht!]
\includegraphics[scale=0.354]{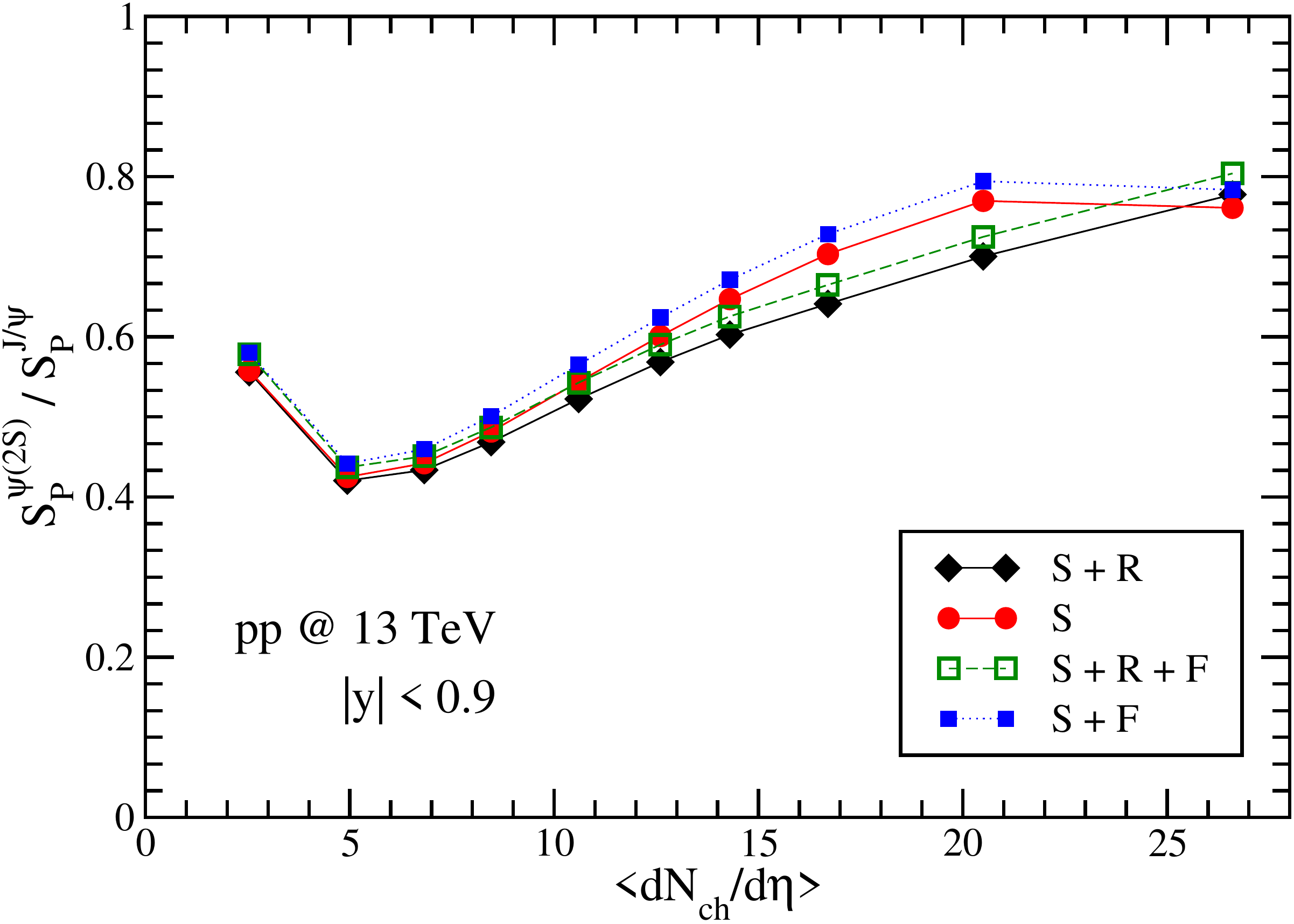}
\caption{The ratio of the survival probabilities of $\psi(2S)$ to $J/\psi$ as a function of final state event multiplicity
for $pp$ collisions at $\sqrt{s} = 13$ TeV.}
\label{c233}
\end{figure}

As a consequence, ``S$+$F'' increases the suppression at high multiplicity, however, an effective regeneration for high resonances neutralizes 
feed-down effect in ``S$+$R$+$F'' (Fig.~\ref{c213}). The suppression ratio plotted in Fig.~\ref{c233} shows that $\psi(2S)$ is 
less suppressed at low and high multiplicities as compared to the intermediate multiplicity bins. 
Because for higher multiplicities, the temperature is  
higher and the regeneration effect creates the difference, which reduces the suppression of $\psi(2S)$ relative
to $J/\psi$.  Results depicted in Fig.~\ref{c233} also show that  individual effect (S)
predicts a slightly more suppression of $\psi(2S)$ than the combined effects (S$+$R) at the highest multiplicity bins.

\subsection{$p_{T}$-dependent Suppression}
Now we discuss the transverse momentum ($p_T$) dependence of charmonia production by considering the $p_T$-dependent survival probability, ($S_P$) for minimum bias ($0-100\%$ centrality) case by taking the weighted average over all centrality bins. 
The weighted average for $S_{P}$ is given by,

\begin{equation}
 S_{P}(p_{T}) = \frac{\sum_{i} S_{P}(p_{T},\langle b_{i} \rangle) W_{i}}{\sum_{i} W_{i}},
 \label{mib}
\end{equation}
where $i = 1, 2, 3, ...$,  indicate various centrality bins corresponding to different multiplicity classes. The weight function $W_{i}$ is  given as
$W_{i} = \int_{b_{i\;min}}^{b_{i\;max}} N_{coll}(b)\pi\; b\; db$. 
The number of binary collision, $N_{coll}$ is calculated by using a Glauber model for $pp$ collisions. In the Glauber model for $pp$ collisions, a azimuthally asymmetric and inhomogeneous density distribution of a proton is considered, which is motivated by the shape of the structure function obtained in deep inelastic scattering. Therefore, same density profile is used to obtained $N_{coll}$ ~\cite{suman}.\\

\begin{figure}[ht!]
\includegraphics[scale = 0.354]{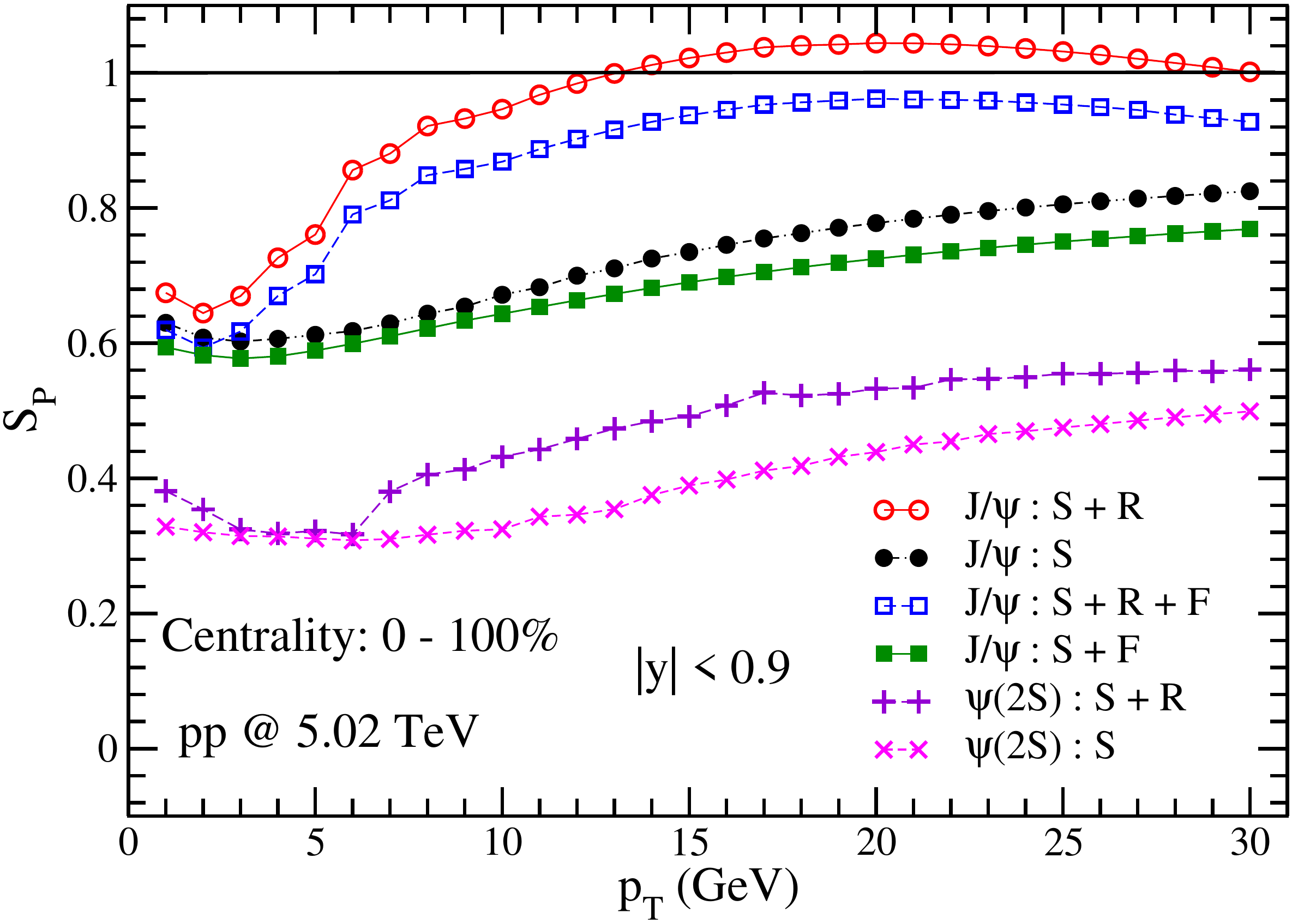}
\caption{Survival probability of $J/\psi$ and $\psi(2S)$ versus $p_{T}$ for $pp$ collisions at $\sqrt{s} = 5.02$ TeV.}
\label{2p11}
\end{figure}

\begin{figure}[ht!]
\includegraphics[scale=0.354]{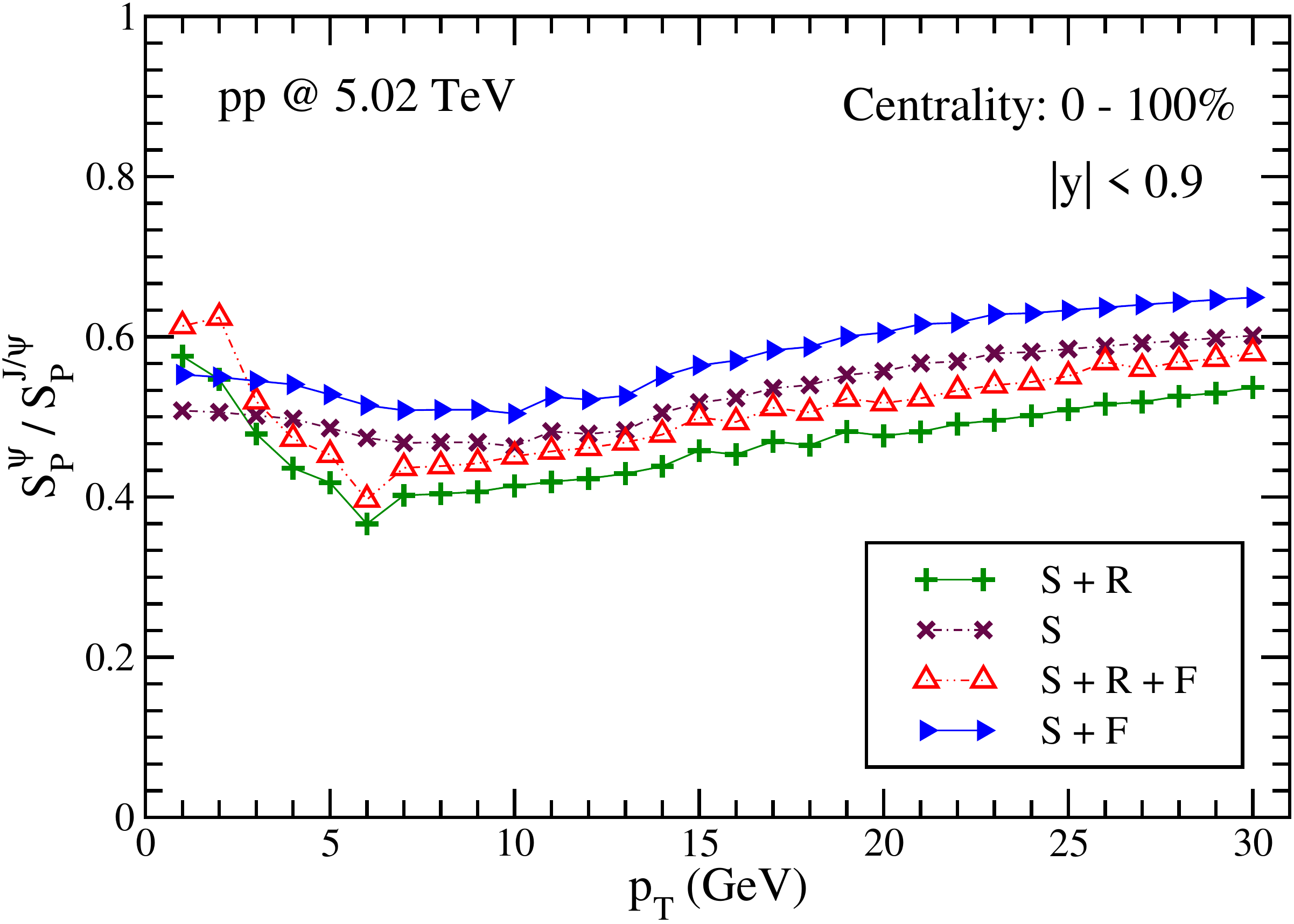}
\caption{The predicted yield ratio  of $\psi(2S)$ to $J/\psi$ for $pp$ collisions at $\sqrt{s} = 5.02$ TeV.}
\label{dbp1}
\end{figure}

The $p_T$-dependent suppression of charmonium is shown in Fig.~\ref{2p11}. 
It shows an increasing trend for regeneration of $J/\psi$ for $p_{T}<20$ GeV and then it slightly decreases
with increasing $p_T$. It seems that the thermodynamic conditions created at $\sqrt{s} = 5.02$ TeV suits the regeneration of $J/\psi$, 
as indicated by the results marked by S$+$R for 
$12 < p_T < 30$ GeV. While including the feed-down along with regeneration (S$+$R$+$F) predicts the suppression 
instead of enhancement for the mentioned $p_{T}$-range. However, for $p_{T}<13$ GeV
both the data sets predict a reduced suppression for $J/\psi$. For $p_{T}<3$ GeV,  $J/\psi$ and $\psi(2S)$ are found to be less 
interactive with the medium, consequently we get a bit less suppression and regeneration at this $p_{T}$-range. 
All the medium effects predict a reduction in suppression with increasing $p_T$. As shown in Fig.~\ref{2p11}, a regeneration effect is observed for 
$\psi(2S)$ for $p_{T} > 6$ GeV. A relative suppression of $\psi(2S)$ over $J/\psi$ is plotted in Fig.~\ref{dbp1}, the double ratio shows that 
inclusion of regeneration (S$+$R) for $\psi(2S)$  reduces its suppression for $p_{T}<3$ GeV. While inclusion of feed-down and exclusion of regeneration 
effect (S$+$F) increases the $J/\psi$ suppression relative to suppression (S) with increasing 
$p_T$. A dip in double ratio at $p_{T} = 6$ GeV, corresponds to the suppression of $\psi(2S)$ when regeneration is 
almost insignificant  while there is a regeneration for $J/\psi$.

\begin{figure}[ht!]
\includegraphics[scale = 0.354]{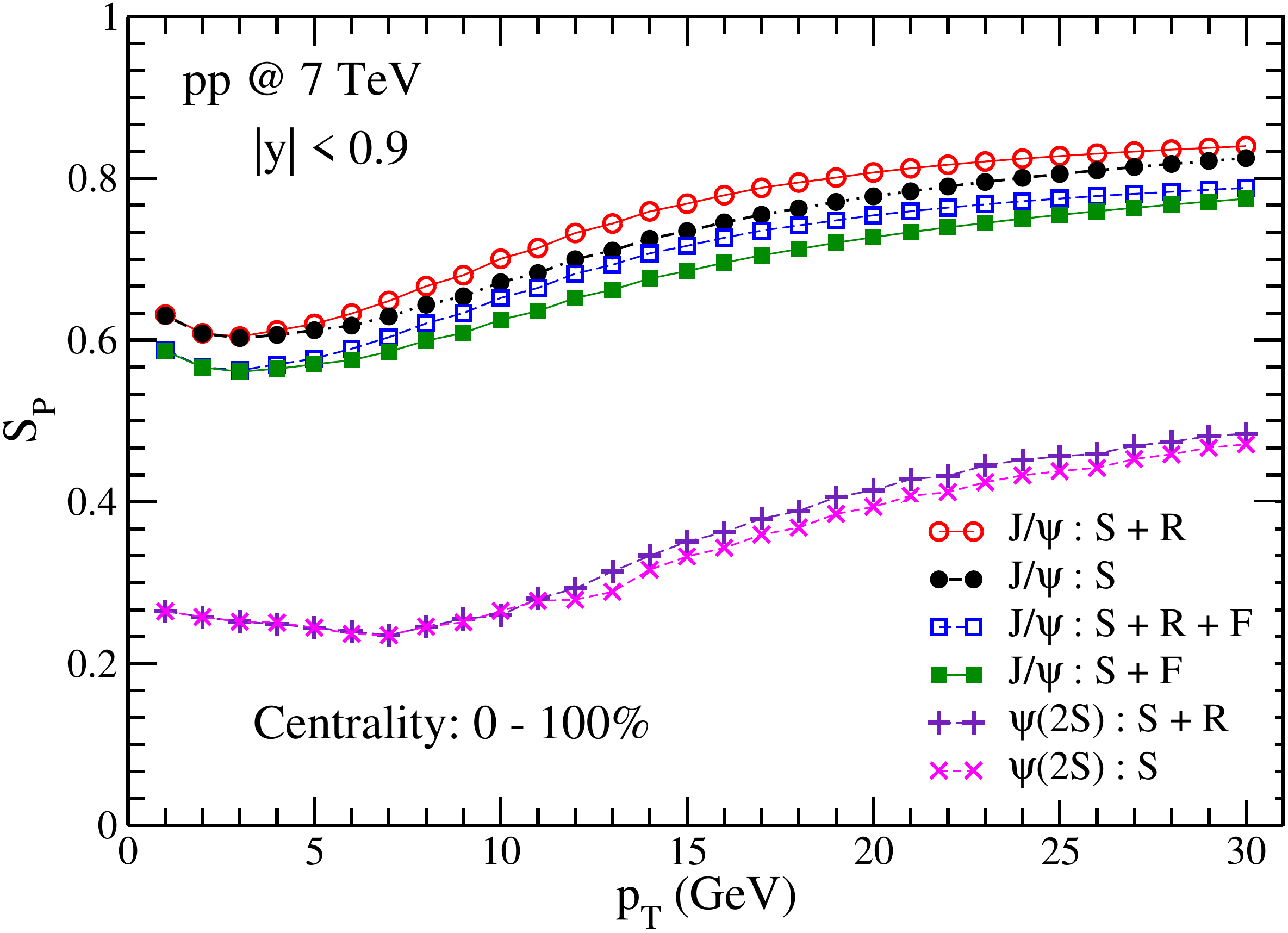}
\caption{Survival probability of $J/\psi$ and $\psi(2S)$ as a function of $p_{T}$ for $pp$ collisions at $\sqrt{s} = 7$ TeV.}
\label{2p12}
\end{figure}

The charmonia suppression for  $\sqrt{s} = 7$ TeV is shown in Fig.~\ref{2p12} as a function of $p_T$.
We find  less suppression at very low $p_{T}$ range: 
$1\le p_T \le 3$ GeV because the multiple 
scattering of the charmonium with medium constituents is less 
effective for very low $p_T$. While for the intermediate $p_T$ range: 
$3\le p_{T} \le 6$ GeV, the scattering with medium will be at the peak, therefore $J/\psi$ receives maximum suppression 
at the intermediate $p_{T}$-range. Further, suppression reduces with increase in $p_T$, because the charmonia 
scattering with medium constituents will be comparably less for high-$p_{T}$ particles as they quickly traverse through the medium
and their abundances are low. 
The regeneration of charmonia increases up to a certain range of the $p_{T}$ and at very high-$p_{T}$ it gets saturated as 
the probability of the recombination of fast moving $c\bar c$ reduces. The $\psi(2S)$ feels maximum suppression for 
intermediate $p_{T}$:  $2\le p_{T} \le 10$ GeV and gets a large suppression than $J/\psi$ due to the low dissociation 
temperature ($T_{D}$) as compared to $J/\psi$ (see Table~\ref{tbII}).\\

\begin{figure}[ht!]
\includegraphics[scale=0.354]{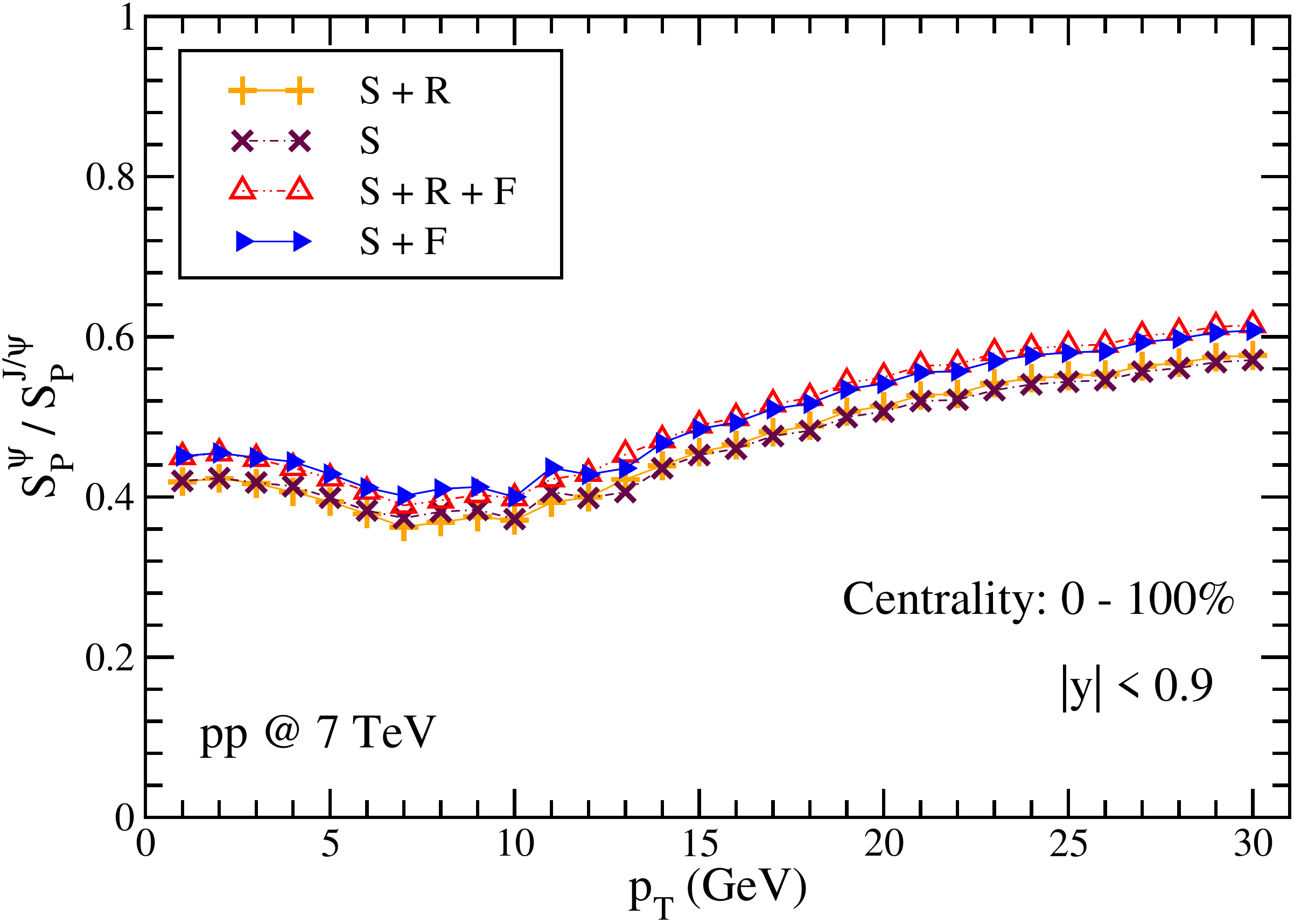}
\caption{The predicted yield ratio of $\psi(2S)$ to $J/\psi$ as a function of $p_{T}$ for $pp$ collisions at $\sqrt{s} = 7$ TeV.}
\label{dbp2}
\end{figure}

The $p_T$-dependent suppression of $\psi(2S)$ over $J/\psi$ corresponding to minimum bias is plotted in 
Fig.~\ref{dbp2}.  It shows that $\psi(2S)$ is  suppressed more over the entire $p_T$-range considered here. 
It  depicts varying  $\psi(2S)$ suppression  from intermediate to high-$p_T$
as the difference in the suppression between $J/\psi$ and $\psi(2S)$ reduces at high-$p_T$ regime. 
The results indicate that the higher resonances are more suppressed than the $J/\psi$ and that is why  the 
inclusion of feed-down correction (S$+$F) always predicts a higher suppression for $J/\psi$.\\
 
\begin{figure}[ht!]
\includegraphics[scale = 0.354]{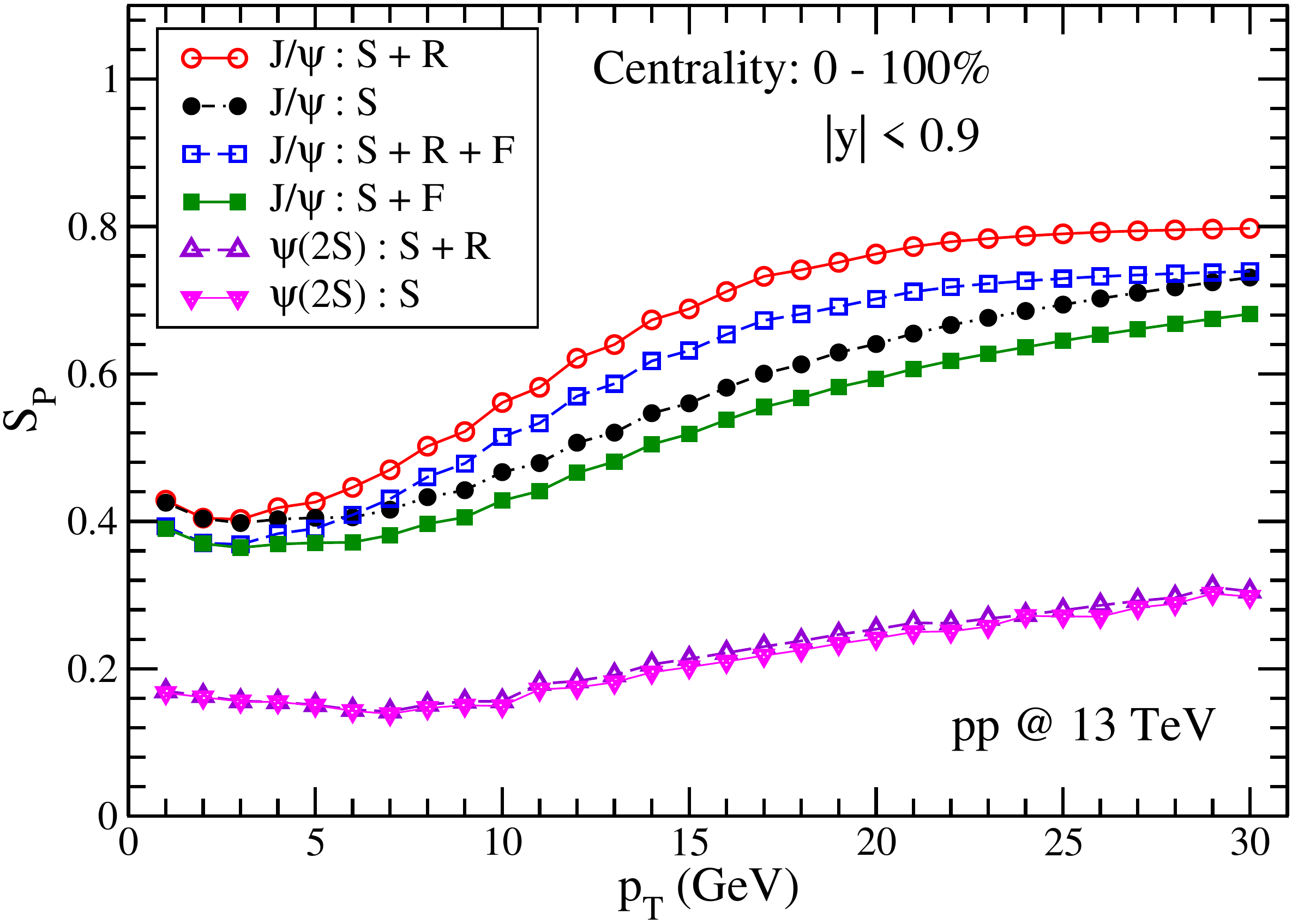}
\caption{Survival probability of $J/\psi$ and $\psi(2S)$ as a function of $p_{T}$ for $pp$ collisions at $\sqrt{s} = 13$ TeV.}
\label{2p13}
\end{figure}

\begin{figure}[ht!]
\includegraphics[scale=0.354]{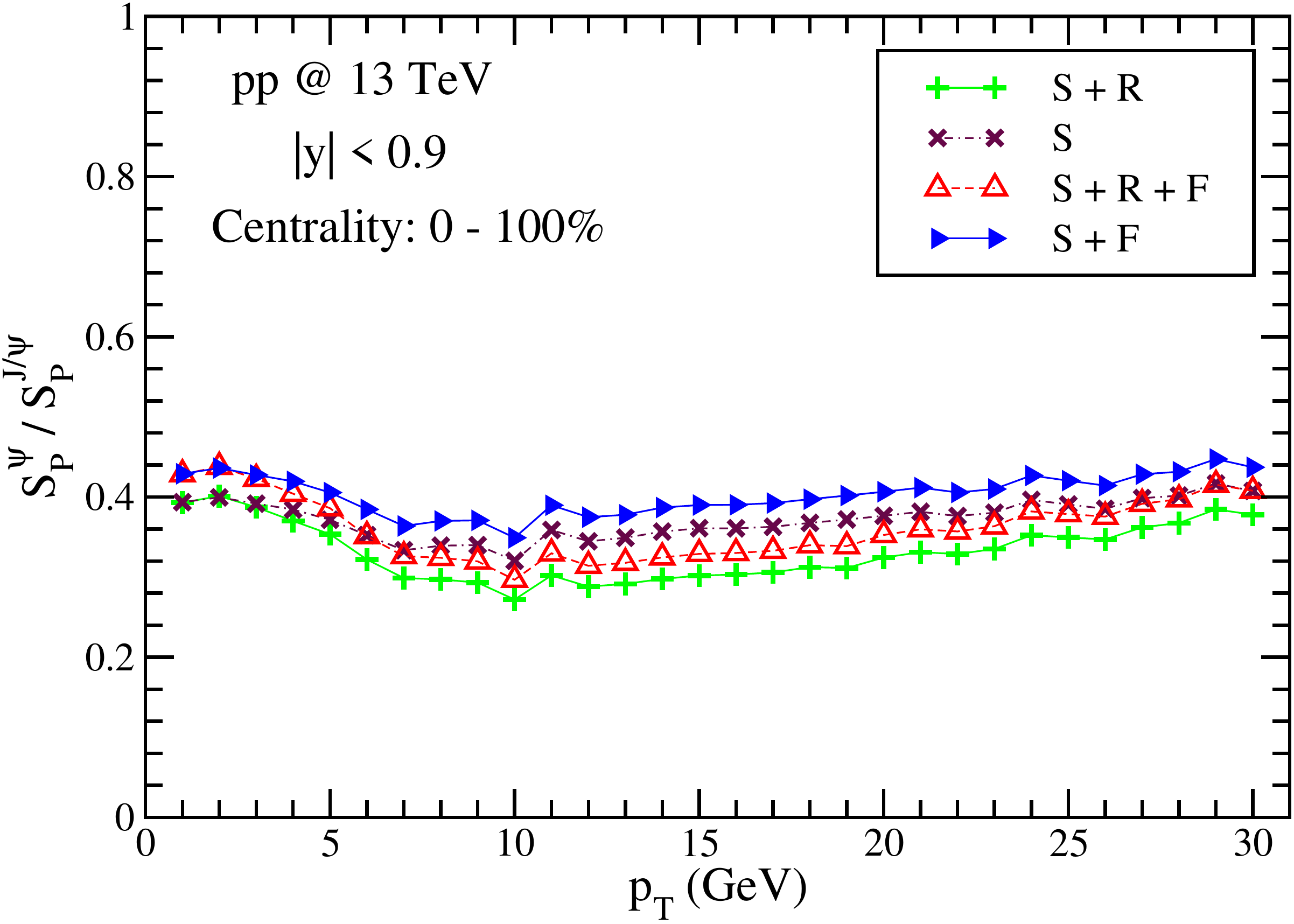}
\caption{The predicted yield ratio  of $\psi(2S)$ to $J/\psi$ is shown for $pp$ collisions at $\sqrt{s} = 13$ TeV.}
\label{dbp3}
\end{figure}

The results displayed in Fig.~\ref{2p13} for $\sqrt{s}=13$ TeV show that the charmonium states are largely suppressed at low-$p_T$ and 
follows a trend similar  to the results obtained for $\sqrt{s} = 7$ TeV.  
But it shows a  larger regeneration for $J/\psi$ for $p_{T}>5$.   The regeneration effect  gets saturated
for $p_T>20$ GeV. The regeneration of $\psi(2S)$ is found to be negligible over the 
chosen $p_{T}$-range, however, 
its suppression reduces at higher transverse momenta: $p_{T}>10$ GeV.
Similar behavior  for double ratio is also obtained in Fig.~\ref{dbp3}, where a large suppression for $J/\psi$ reduces the relative suppression 
for $\psi(2S)$, i.e. ``S$+$F''. However, the relative suppression of $\psi(2S)$ over $J/\psi$ comes
out to be approximately constant, a slightly reduced suppression may be observed at high-$p_{T}$ regime (Fig.~\ref{dbp3}).\\

It is worthwhile to mention that in this work an interplay between regeneration and feed-down mechanisms for $\sqrt{s} =$ 5.02 to 13 TeV
has been demonstrated. In UMQS, not a single parameter is freely varied in order to explain the experimental data or predict the charmonium
suppression.

\section{Conclusions}
The present analysis shows a QGP-like medium formation for all the multiplicity bins in ultra-relativistic $pp$ collisions at LHC energies. 
We have observed that $\psi(2S)$ is suppressed more and the effect of regeneration is also marginal for $\psi(2S)$ 
as compared to $J/\psi$. This makes $\psi(2S)$ a relatively cleaner probe to investigate QGP-like medium in the small collision systems.
As the cold nuclear environment is absent in the $pp$ collision systems any modification in the charmonium yield can be considered as a pure 
effect of hot QCD medium.  In contrast to this, the system formed in $pA$ collision will show the combined effects of the hot and cold nuclear
matter, therefore, $pp$ collisions can be used as a scale to distinguish  CNM effects. The $pp$ collision
results which are routinely used as a benchmark to understand the AA collision system is now become questionable because
of the possibility of the creation of QGP-like medium in $pp$ interactions.
Thus, the formation of QGP in $pp$ collisions will create a new puzzle and inevitably a new baseline will be required 
to analyze AA collisions.  

\noindent
\section{Acknowledgments}

This research work has been carried out under financial supports from DAE-BRNS, Government of India, Project No. 58/14/29/2019-BRNS of Raghunath Sahoo. CRS and 
RS acknowledge the financial support 
under the above BRNS project. Further R.S. acknowledges the financial supports under the CERN Scientific Associateship, CERN, Geneva, Switzerland. The authors would like to acknowledge interesting discussions with Dr. Sushanta Tripathy. The authors acknowledge the Tier-3 computing facility in the experimental high-energy physics laboratory of IIT Indore supported by the ALICE project.

\section{Appendix}
\begin{figure*}[h!]
 \includegraphics[scale=0.445]{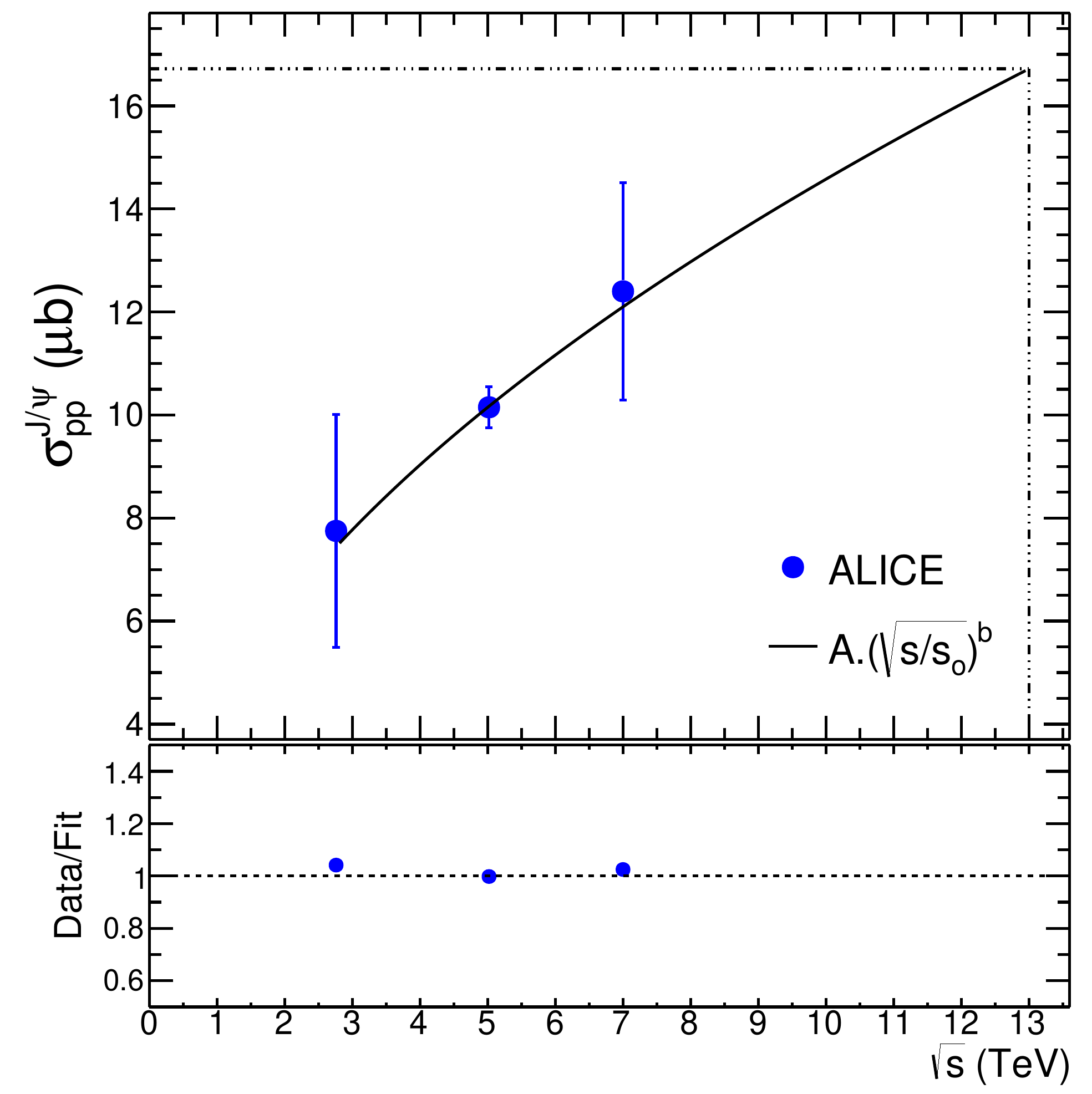}
  \includegraphics[scale=0.445]{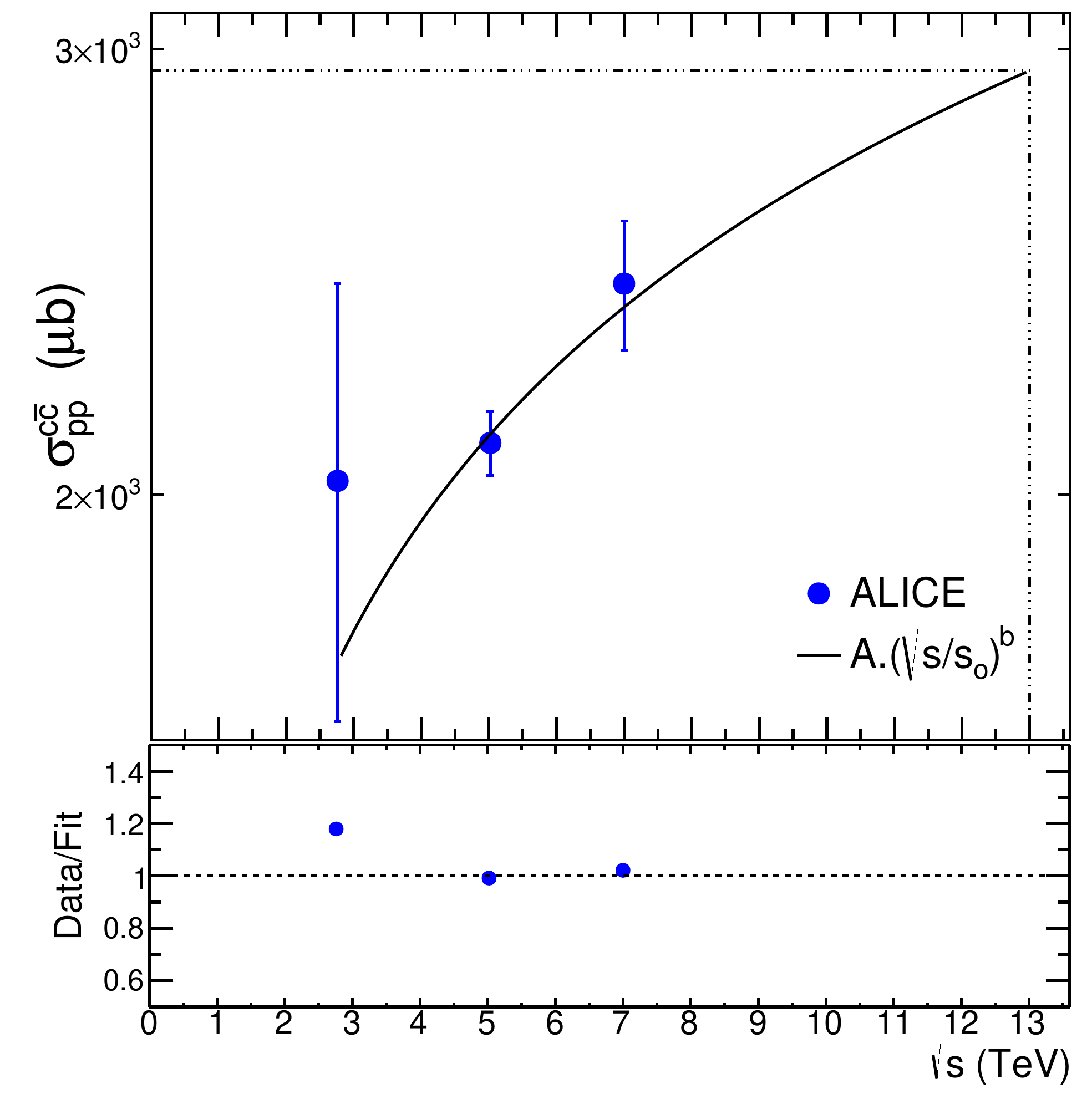}
 \caption  {(Color online) Top panel shows the energy dependence of inclusive J/$\psi$ (left) and $c\bar{c}$ (right) production cross-section at mid-rapidity in $pp$ collisions as fitted by a power-law function. The solid blue circles are ALICE data~\cite{ALICE:2021dhb,ALICE:2012vup,ALICE:2019pid,ALICE:2011zqe} and the black curve represents the power-law function. The quadratic sum of statistical and systematic uncertainties of ALICE data are presented in a single error bar. Bottom panels show the ratio between ALICE data and the fit. The dashed lines shows the value of  $\sigma_{J/\psi}^{pp}$ (top left) and $ \sigma_{c\bar{c}}^{pp}$ (top right) corresponding to $\sqrt{s} = 13$ TeV. }
 \label{fig:extrapolation}  
 \end{figure*}

To extrapolate the energy dependence of inclusive J/$\psi$ and $c\bar{c}$ production 
cross-section at mid-rapidity in $pp$ collisions, we have adopted the method of functional
form fitting as discussed in Ref.~\cite{Bossu:2011qe}. In this method, the  
energy dependence of the inclusive J/$\psi$ and $c\bar{c}$ 
cross section are fitted with a power-law distribution of the form:
\begin{equation}
f(\sqrt{s}) = A\cdot \left(\sqrt{s/s_0}\right)^b,
\end{equation}
where $A$, $s_0$ and $b$ are free parameters. It is worth noting here that although it is not derived from a first principle, such a method of extrapolation reproduces the measured distribution very well.
In  Fig.\ref{fig:extrapolation} the solid blue circles are ALICE data and the black curves represent fitting function. The quadratic sum of statistical and systematic uncertainties of ALICE data are presented by a single error bar. The results obtained by extrapolation at $\sqrt{s}$ = 13 TeV for inclusive J/$\psi$ and $c\bar{c}$ production cross-sections are reported in Table~\ref{tbI}. Recent ALICE result~\cite{ALICE:2021dtt} shows that the J/$\psi$ production cross-section at mid-rapidity is $\approx$ 16.15 $\pm$ 0.54 $\mu$b (quadratic sum of statistical and systematic uncertainties). This value is very close to our extrapolated result of 16.72 $\mu$b. This agreement provides a strong case for our extrapolated result of $c\bar{c}$ production cross-section.

 \end{document}